\documentstyle[11pt,aaspp4,epsfig]{article}

\newcommand{\et}{et\thinspace al.\ }
\newcommand{\bvr}{$BV\!R$}

\newcommand{\bmv}{$B$$-$$V$}
\newcommand{\vmr}{$V$$-$$R$}
\newcommand{\bmr}{$B$$-$$R$}
\newcommand{\cmt}{$C$$-$$T_1$}
\slugcomment{}

\lefthead{Rhode \& Zepf}
\righthead{Globular Cluster System of NGC~4472}

\begin{document}

\title{The Globular Cluster System in the Outer Regions of NGC~4472}

\author{Katherine L. Rhode\altaffilmark{1,2} and Stephen E. Zepf\altaffilmark{1}} 
\affil{Department of Astronomy, Yale University, New Haven, CT 06520;\\ rhode@astro.yale.edu, zepf@astro.yale.edu}

\altaffiltext{1}{Visiting Astronomer, Kitt Peak National Observatory,
National Optical Astronomy Observatories, which is operated by the
Association of Universities for Research in Astronomy, Inc. (AURA)
under cooperative agreement with the National Science Foundation}
\altaffiltext{2}{NASA Graduate Student Researchers Program Fellow}

\begin{abstract}

We have undertaken a study of the globular cluster (GC) systems of a
large sample of elliptical and spiral galaxies in order to test
specific predictions of galaxy formation models.  Here we present
results for the first elliptical in the study, the giant Virgo cluster
galaxy NGC~4472 (M49).  The galaxy was observed in three filters
(\bvr) using the wide-field Mosaic Imager on the Kitt Peak 4-meter
telescope.  The Mosaic images roughly double the radial coverage of
previous CCD observations of NGC~4472.  We have combined the Mosaic
data with published spectroscopic data and archival HST observations
in order to study NGC~4472's GC system in detail, and to fully
characterize the amount of contamination in our sample of GC
candidates.  We find that the radial profile of the GC system is fit
fairly well by a de~Vaucouleurs law of the form log~$\sigma_{GC}$ =
(3.38$\pm$0.08) $-$ (1.56$\pm$0.05) $r^{1/4}$ out to 17\arcmin\
($\sim$80 kpc), but that the observed profile falls slightly below the
de~Vaucouleurs law between 17\arcmin\ and 23\arcmin, the limit of the
data.  The bimodal color distribution observed in previous studies is
apparent in our data.  We find a small metallicity gradient in the
inner 8\arcmin\ of the total GC system due to the increasing ratio of
blue to red clusters, consistent with results from past studies.  The
gradient vanishes, however, when the entire radial extent of the data
is taken into account.  We estimate a total of $\sim$5,900 GCs in
NGC~4472 out to 23\arcmin, yielding a specific frequency of
3.6$\pm$0.6 within this radius. This specific frequency value is
smaller than that found by previous studies of NGC~4472.  We examine
the implications of these results with regard to predictions made by
four different galaxy formation models, and find that all four models
have at least one inconsistency with the data.
\end{abstract}

\keywords{galaxies: formation; galaxies: star clusters; galaxies:
individual (NGC~4472)}

\clearpage
\section{Introduction}

Globular clusters (GCs) are among the oldest readily-observable
objects in the Universe, and have been detected in all types of
massive galaxies and many dwarfs.  The fact that GCs have such large
ages and are seen in virtually all types of galaxies indicates that
they are closely tied to galaxy formation.  Studying the ensemble
properties of the entire system of GCs around a given galaxy can
provide important clues regarding the formation and subsequent
evolution of that galaxy, and such studies can be particularly useful
for testing theories of elliptical galaxy formation.  GC systems of
ellipticals are typically more populous than those of spirals (cf.\
\cite{az98}).  Moreover, GC systems are generally at least as extended
as the light distributions of their host galaxies (cf.\
\cite{harris86}, \cite{az98}), so GCs can be used to probe the
formation and enrichment history of the outer regions of elliptical
galaxies.

Recent studies have found that the GC systems of many ellipticals have
bimodal rather than single-peaked color distributions (e.g.,
\cite{geisleretal96}, \cite{kunwhit99}, \cite{gebhardt99}).  The
presence of two peaks has been interpreted as due to two GC
populations with different metallicities, possibly formed in different
episodes.  Evidence also suggests that at least in some ellipticals,
the bluer (metal-poor) GC population is more extended than the red
(metal-rich) population (e.g., \cite{geisleretal96},
\cite{neilsen00}).  These observations agree in general with the
predictions of a galaxy formation scenario proposed by Ashman \& Zepf
(1992), in which an elliptical galaxy is formed via the merger of two
or more disk galaxies.  In the Ashman \& Zepf picture, a younger,
spatially-concentrated, metal-rich system of GCs is formed during the
merger itself, and the spatially-extended, metal-poor GC population
originates in the halos of the progenitor spirals.  Other models may
also explain the observed differences between the GC populations in
ellipticals.  Forbes \et (1997) suggest that the GCs in ellipticals
are formed in distinct episodes, with the metal-poor population formed
early, from less enriched gas, and the metal-rich population forming
later, along with the galaxy field stars.  In the case of giant
ellipticals and cD galaxies, Forbes \et suggest that tidal stripping
of GCs from nearby galaxies may contribute to building up the outer GC
system.  Another model, proposed by C\^ot\'e \et (1998), suggests that
giant ellipticals form via dissipational collapse, producing a
metal-rich GC population.  Metal-poor GCs are then captured from other
galaxies by accretion or tidal stripping.

These scenarios make different predictions for the detailed properties
of the GC systems of ellipticals.  For example, the merger scenario of
Ashman \& Zepf predicts that the mass-normalized number of metal-poor
GCs around ellipticals and spirals should be the same, since in this
model all of the metal-poor GCs come from the progenitor spirals.
The C\^ot\'e \et model predicts that a significant number of
metal-poor, tidally-stripped GCs should reside at large distances
from giant cluster ellipticals.  We have undertaken a study of a large
sample of elliptical and spiral galaxies in order to test the
predictions of the various formation models.  Using a new generation
of wide-field CCD imagers, we have observed a sample of early-type
galaxies in the Virgo cluster and the field, as well as a large sample
of fairly isolated, edge-on spirals.  Our aim is to accurately
quantify the total numbers, spatial distributions and color
distributions of the GC systems in these galaxies out to large radii.
By observing the galaxies in three filters and employing image
classification analysis we can greatly reduce the amount of potential
contamination (from background and foreground objects) in our GC
samples, while simultaneously detecting GCs further out than previous
CCD studies have done.  Furthermore, we make use of available
space-based and spectroscopic data to help quantify the amount of
contamination that exists in our GC lists.

Here we present the results of our study of the GC system of the
galaxy NGC~4472 (M49). NGC~4472 is located four degrees from the
center of the Virgo cluster and is the brightest of the Virgo
galaxies.  We chose to analyze NGC~4472's GC system first because the
abundance of Hubble Space Telescope (HST) and spectroscopic data
currently available made it possible for us to fully quantify the
amount of contamination in our data.  Studying this galaxy has allowed
us to test our overall methods and ensure that they are sufficient for
accomplishing the goals of the project.  In the following section, the
observations and initial data reduction are described.
Section~\ref{section:analysis} describes subsequent analysis steps
performed on the data, and Section~\ref{section:results} gives the
results.  Section~\ref{section:models} consists of a discussion of the
implications of the results with regard to various galaxy formation
models, and the final section contains a summary of the paper.

\section{Observations and Reductions}
\label{section:observations}

Observations of NGC~4472 were obtained in March 1999 with the Mosaic
Imager on the 4-meter Mayall telescope at Kitt Peak National
Observatory (KPNO). The Mosaic Imager consists of eight 2048 x 4096
SITe CCDs.  The CCDs are separated by small ($\sim$50 pixels) gaps.
When mounted on the 4-meter, each pixel of the Mosaic subtends
0.26\arcsec\ on the sky, resulting in a total field-of-view 36\arcmin\
on a side.  We imaged the galaxy in three broadband filters (\bvr).
Five exposures were taken through each filter, and the telescope was
dithered in a predetermined pattern on the sky, in order to help
eliminate cosmic rays as well as blank regions in the images due to
the gaps between the CCDs.  Total integration times were 3300 s in
$B$, 2700 s in $V$, and 2100 s in $R$.  The $V$ and $R$ images were
taken under fairly clear, although not photometric, conditions, but
the $B$ image was taken under poorer sky conditions.  Observations in
three filters were taken of three more early-type galaxies (NGC~3379,
NGC~4406, and NGC~4594) during the same observing run; results for
these objects will be presented in a subsequent paper (\cite{rz01}).

Data reduction was accomplished using the IRAF\footnote{IRAF is
distributed by the National Optical Astronomy Observatories, which are
operated by the Association of Universities for Research in Astronomy,
Inc., under cooperative agreement with the National Science
Foundation.} package MSCRED, which is designed especially for the
reduction of images from mosaic CCD detectors.  We reduced the data
using standard practices, i.e., ZEROCOMBINE to create a combined bias
image, FLATCOMBINE and SFLATCOMBINE to create combined dome flats and
twilight sky flats, MEDIAN to smooth the sky flats, and CCDPROC to
perform overscan level subtraction, bias image subtraction, and
flat-field division on the object images.  We then used the MSCRED
tasks MSCZERO, MSCCMATCH and MSCIMAGE to construct single images from
the multi-extension mosaic images, by mapping the pixels from each
FITS extension to a single grid on the sky.  A constant sky background
was subtracted from the resultant individual images using the task
IMSURFIT.  Finally, multiple images in a given filter were scaled to a
common flux level and combined into a single ``stacked'' image using
pixel rejection to eliminate cosmic rays.  A constant sky level,
measured from the image used as a reference image during the scaling
step, was added to each of the stacked images to restore the
background counts.  The resolution (point-spread function FWHM) of the
final stacked images is 1.1\arcsec\ in $B$, 1.1\arcsec\ in $V$, and
1.3\arcsec\ in $R$.

\section{Data Analysis}
\label{section:analysis}

\subsection{Detection of the GC System}

The primary observational signature of a GC system around a distant
galaxy is an overdensity of compact objects around the galaxy with
magnitudes and colors in the expected range for GCs.  At a distance
of $\sim$16 Mpc, GCs around NGC~4472 are unresolved in our images.
To detect the GCs, we began by smoothing each of the three stacked
images using a circular filter with a diameter 6 times the average
FWHM in the image.  The light from the galaxy was then removed by
subtracting this smoothed image from the original.  A constant was
added to the resultant images to restore the background level, and the
\bvr\ images were aligned to each other.  Next the IRAF task DAOFIND
was used to search for GC candidates in the images.  A few small
regions of the images --- e.g., around saturated stars, or on the
edges of the frames, where the noise level was higher --- were
excluded from the search because DAOFIND tended to find spurious
sources in those regions.  A list of 4,329 GC candidates was produced
by requiring that the candidates be detected in all three filters.

\subsection{Extended Source Cut}

To eliminate contaminating objects (specifically, background galaxies)
from the list of GC candidates, the FWHM values of the light profiles
of the 4,329 sources were measured and plotted versus their
instrumental magnitudes.  Objects with FWHM values that were much
larger than the average for the image were presumed to be extended
objects and were removed from the GC candidate list.

\begin{figure}
\plotone{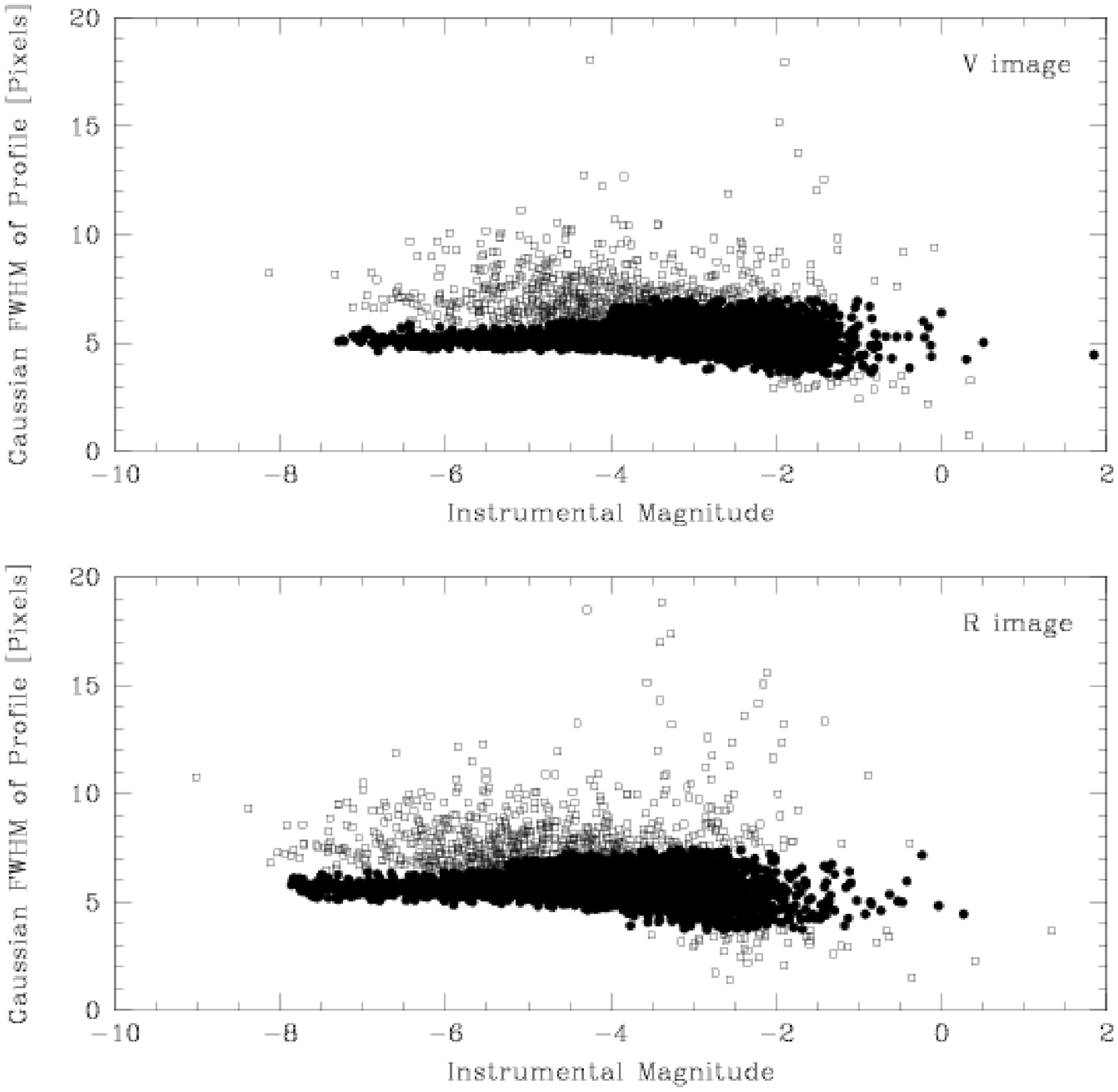} 
\caption{{\bf Extended source cut.} Gaussian FWHM of the radial
profile versus instrumental magnitude for the 4,329 sources in the
initial GC candidate list in the $V$- and $R$-band images.  Filled
circles are objects that pass the FWHM cut; open squares are objects
eliminated from the GC candidate list at this step.
\protect\label{fig:fwhm mag}}
\end{figure}

Figure~\ref{fig:fwhm mag} shows the FWHM versus magnitude values for
the 4,329 GC candidates in the $V$ and $R$ images.  Filled circles
mark the accepted objects, and open squares are objects thought to be
extended and therefore eliminated from the GC list.  The figure shows
that at bright magnitudes, point sources tend to have FWHM values that
scatter around a mean value and form a distinct sequence in the FWHM
versus magnitude plane.  As the signal-to-noise ratio decreases at
fainter magnitudes, however, the sequence spreads out and it becomes
more difficult to distinguish between point sources and extended
objects.  We took this into account by accepting a larger range of
FWHM values for sources with fainter magnitudes.  Sources that had
FWHM values too small to be believable were also removed from the GC
candidate list in this process.  By checking the results of our
extended source cut against HST WFPC2 images of NGC~4472 (see
Section~\ref{section:HST data}), we found that the best results were
obtained by basing the extended source cut on the $V$ and $R$ images
only, and excluding the $B$ image (which was taken under deteriorating
sky conditions) from this process.  871 sources that appeared extended
in {\it either} the $V$ or $R$ image were eliminated in the extended
source cut, leaving 3,458 objects in the GC candidate list.

\subsection{Photometry and Photometric Calibration}

Because the GC candidates are faint, the most accurate photometry can
be achieved by measuring their light in a small aperture and applying
an aperture correction to account for the light missing from the part
of the profile that has been excluded.  We determined an aperture
correction for each of the three stacked images by measuring
$\sim$20$-$30 bright, unsaturated stars in each image and measuring
their magnitudes in a series of apertures from 1$-$6 times the average
FWHM in the image.  The measured aperture corrections and associated
errors are given in Table~\ref{table:apcorr}.  The errors range from
0.01 to 0.02 magnitudes, and reflect the fact that the point spread
function varies slightly across a given Mosaic frame.  Using a single
aperture correction for a given stacked image limits the accuracy of
the final photometry to 0.01 to 0.02 magnitudes.

The 4-m observations of NGC~4472 were taken during non-photometric
conditions, so in order to calibrate them, we observed the galaxy with
the WIYN\footnote{The WIYN Observatory is a joint facility of the
University of Wisconsin-Madison, Indiana University, Yale University,
and the National Optical Astronomy Observatories.} 3.5-m telescope at
KPNO on a photometric night in January 2000.  One 600 s exposure of
NGC~4472 was taken through each filter (\bvr).  Calibration frames
were taken throughout the night.  A total of nine standard stars
(\cite{land92}) were used to calibrate the WIYN images.  Twenty-three
bright, but not saturated, stars were found in common between the WIYN
and 4-m images and used to ``bootstrap'' the 4-m calibration to the
WIYN calibration.  The WIYN images were essentially treated as
standard star fields; magnitudes and colors of the 23 bright stars
were measured in both images, and photometric coefficients (zero
points and color terms) were derived to calibrate the 4-m data to the
WIYN data.

Finally, aperture photometry was carried out in all three filters on
the 3,458 GC candidates in the Mosaic frames, using an aperture of
radius equal to the average FWHM of each image.  Calibrated total
magnitudes for the sources were calculated using the aperture
corrections listed in Table~\ref{table:apcorr} and the photometric
coefficients derived via the bootstrapping method.  The magnitudes
were corrected for Galactic extinction using reddening maps from
Schlegel \et (1998), which give values of $A_B =$ 0.087, $A_V =$
0.067, and $A_R =$ 0.057 in the direction of NGC~4472.

To check the accuracy of our photometric calibration, we compared the
$R$ magnitudes of 20 of our GC candidates with magnitudes from a study
of NGC~4472's GC system published by Geisler \et (1996).  Because
Geisler \et did not publish equatorial positions for their GCs, we
used a list of objects published in Sharples \et (1998), who obtained
spectra for 55 of the GC candidates from the Geisler \et study.  The
Sharples \et paper tabulates precise coordinates for the sources, and
lists the $T_1$ and \cmt\ values from Geisler \et We used the
coordinates of the objects to find matches between their list of GC
candidates and ours, then used a transformation given in Geisler
(1996) to convert the $T_1$ magnitudes to $R$.  The mean difference
between our $R$ magnitudes and those calculated from the Geisler \et
data is 0.005, with a standard deviation of 0.035.  Geisler \et
estimate their photometric errors to be $\sim$0.03 magnitudes, and
ours are comparable.  This indicates excellent agreement between our
photometry and that from Geisler \et

\subsection{Color Selection}
\label{section:color cut}
The fact that we observe our target galaxies with three broadband
filters allows us to significantly reduce the amount of contamination
in our GC samples, because we can then use two colors to differentiate
the GCs from other stellar populations or background objects. Before
beginning this project, we investigated which color combinations would
help us to adequately isolate the GCs in the color-color plane.  We
found that by using \bvr\ filters, we could distinguish GCs from the
majority of other types of objects, while not having to use
impractically long exposure times.  Figure~2$a$ shows where GCs would
be expected to lie in the \bvr\ color-color plane.  The large filled
circles are Milky Way GCs from the Harris catalog (\cite{harris96}).
We used the Harris catalog to derive a linear relationship between
\bmv\ and \vmr\ color for Milky Way GCs.  This relationship serves as
the mid-line of the rectangular box shown.  The mid-line extends from
\bmv\ $=$ 0.56 to 0.99, which corresponds to an [Fe/H] range of $-$2.5
to 0.0 for Galactic GCs.  The rectangle is drawn by extending the
fitted \vmr\ versus \bmv\ relationship by $\pm$2 times the RMS scatter
about the fit.  The positions of galaxies of different morphological
types are shown in the figure as ``tracks'' the galaxies would follow
with increasing redshift.  The tracks were made using template spectra
from Kennicutt (1992), Coleman, Wu \& Weedman (1980), and Kinney et
al. (1996).  The spectra were redshifted from z$=$0 to z$=$0.7 in
steps of 0.025, and then sampled through the \bvr\ filter transmission
curves to arrive at the data plotted in the figure.

Figure~2$a$ shows that GCs should occupy a fairly narrow region in the
\bvr\ color-color plane, and also that the galaxies that have colors
similar to GCs are, for the most part, late-type objects with low to
moderate redshifts.  Many of these background galaxies can be
eliminated with image classification analysis, but others would be
impossible to eliminate using only ground-based photometric data.  For
this reason we have used space-based and spectroscopic data to help us
quantify the amount of contamination in the sample from such objects
(see Section~\ref{section:contamination}).

To eliminate non-GCs from our list of 3,458 GC candidates, we devised
a selection algorithm that works as follows: (1) If a GC candidate has
a V magnitude brighter than 19.5 (the brightest magnitude for GCs in
the Virgo elliptical M87 (\cite{whitmore95})), it is removed from the
list.  (2) If the GC candidate has \bmv\ and \vmr\ colors that place
it within the rectangular region shown in Figure~2$a$, or within a
distance from the rectangle equal to its formal photometric error, it
is retained in the sample.  A total of 1,465 objects pass the
magnitude and color criteria imposed on the GC candidate list; these
objects serve as the final sample of GC candidates.  The positions of
the accepted objects in the \bvr\ color-color plane are shown in
Figure~2$b$.  Because the photometric errors for these objects were
taken into account in the color selection step, they occupy a region
broader than the rectangular area shown in Figure~2$a$.

\subsection{Completeness Tests}
\label{section:completeness}

To investigate the GC detection limit in our images, we ran a series
of tests of the completeness in each filter.  For these tests, 800
artificial point sources of known magnitude were added to each image,
covering a range of 5$-$7 magnitudes at 0.1-magnitude intervals.  The
same detection steps used on the original image (i.e., DAOFIND with a
given detection threshold) were applied to the images containing the
artificial sources, to see what fraction of the objects in each
magnitude interval would be detected.  The results from these tests
yield the completeness as a function of magnitude for each filter;
these completeness functions are shown in
Figure~\ref{fig:completeness}.  The 50\% completeness limits in the
three filters are: $B$ = 23.33, $V$ = 23.65, and $R$ = 22.83.

The $B$ image was taken under poor sky conditions and is therefore
much more shallow than intended. This has serious consequences for the
completeness of our GC sample as a whole.  Because typical \bmv\
colors for GCs in ellipticals lie in the range 0.55 $-$ 1.0, the $B$
image must be that much deeper than the $V$ image in order for all of
the GCs that are detected in $V$ to be detected in $B$ as well.  Our
$B$ image is in fact slightly more shallow than the $V$ image.  The
implications of this are that in our overall sample of GC candidates,
we are better able to detect blue GCs (since redder GCs will not
show up as well in our relatively shallow $B$ image), and also that
the comparatively poor completeness of the $B$ image will dominate the
completeness corrections we make in later steps (e.g., in the
correction for magnitude incompleteness made in the calculation of the
total number of GCs; see Sections ~\ref{section:GCLF} and
~\ref{section:spec freq}).

\subsection{Contamination}
\label{section:contamination}
For ground-based, photometric studies of extragalactic GC systems, one
of the most serious potential problems is contamination of the GC
candidate sample by foreground and background objects.  By selecting
GC candidates using image analysis and color selection in three
filters, contamination can in principle be greatly reduced, although
not eliminated completely.  We used space-based and ground-based data,
as well as theoretical models, to quantify the amount of contamination
existing in our GC sample.

\subsubsection{HST Data}
\label{section:HST data}

The Hubble Space Telescope has the advantage of being able to resolve
many faint compact background galaxies that might appear as point
sources in ground-based images and could contaminate a GC sample.  To
make use of this fact, we analyzed all of the publicly-available
Wide-Field and Planetary Camera 2 (WFPC2) images\footnote{Based on
observations made with the NASA/ESA Hubble Space Telescope, obtained
from the data archive at the Space Telescope Science Institute.  STScI
is operated by the Association of Universities for Research in
Astronomy, Inc. under NASA contract NAS 5-26555.} of NGC~4472 taken
with the F555W filter, the filter that most closely resembles the $V$
filter.  There were four data sets that fit these criteria: HST
program GO.5236, PI: Westphal; HST program GO.6673, PI: Baum; HST
program GO.5920, PI: Brodie; and HST program GO.6352, PI: Ferguson.
We requested that ``On-The-Fly'' calibration be applied to the images
returned by the HST data archive.

Images of a given pointing that had been dithered were shifted so that
they were aligned to each other.  The CRREJ task, in the STSDAS
package, was then used to create combined, cosmic-ray-rejected images
of each pointing in the four data sets.  A few of the pointings
slightly overlapped each other.  We eliminated three small sections of
images from the rest of the analysis in order to avoid duplicating the
coverage of any area on the sky.  Excluding the redundant regions, the
total area around NGC~4472 covered by the HST images is 22.05 square
arc minutes.  The IRAF task WCSCTRAN was used to translate the sky
positions of the 1,465 GC candidates into pixel positions in the WFPC2
images.  A total of 133 of these sources were located in the WFPC2
images, distributed over two planetary camera (PC) fields and 12 Wide
Field camera (WF) fields.

We followed the method used by Kundu \et (1999) to distinguish between
point sources ({\it bona fide} GC candidates) and extended objects
(contaminating background galaxies) in the WFPC2 images.  Aperture
photometry was performed on the 133 sources using aperture radii of
0.5 pixels and 3 pixels and a sky annulus from 5 to 8 pixels.  Kundu
\et found that the ratio of fluxes in the two apertures is a good
criterion for eliminating background galaxies: objects appearing on
the PC chip with Counts$_{3pix}$/Counts$_{0.5pix}$ $<$ 12, and on the
WF chips with Counts$_{3pix}$/Counts$_{0.5pix}$ $<$ 8, are point
sources and thus possible GCs.  Using these criteria, we found that
five of the 133 GC candidates in the WFPC2 images were actually
extended objects.

Besides investigating how many of the GC candidates were likely
background galaxies, we also wanted to make use of HST's ability to
distinguish between point-like and extended objects in order to
fine-tune the extended source cut we had applied early in the GC
detection process.  We found that by using all three of the Mosaic
images ($B$, $V$, and $R$) to apply the extended source cut, we
eliminated some objects that, according to the HST data, were point
sources.  These objects appeared slightly ``blurred out'' in the $B$
image because of its relatively poor quality, but were not extended in
the $V$ and $R$ Mosaic images.  The best results (i.e., most
consistent with the results from the HST data) were obtained when we
used only the $V$ and $R$ images to apply the extended source cut.

Even with our fine-tuned extended source cut, at faint magnitudes we
are still throwing out a few objects that are not intrinsically
extended but may appear so because their profiles are just above the
noise level of the image.  We investigated the profiles of a sample of
sources that were excluded in the extended source cut but had colors
in the range for GCs.  We found that approximately 3.6\% of the GC
candidates with the correct colors are mistakenly eliminated in the
extended source cut.  Both this effect and the contamination of the GC
sample are minimized if the sample is cut at $V$ $\leq$ 23.5,
indicating the importance of signal-to-noise for reliable image
classification.  Applying this magnitude limit to the GC candidate
list leaves a total of 1,461 objects.  For this sample, there are four
contaminating objects out of a total of 131 GC candidates appearing in
the HST images. The contamination can be characterized in terms of a
surface density of contaminating objects for the sample of 1,461 GC
candidates; there are 4 contaminating objects found in the 22.05
square arc minutes covered by the WFPC2 data, or 0.181 objects per
square arc~minute.  Note that this is an estimate of the amount of
contamination existing in our sample from background galaxies, and
does not include contamination from foreground stars.  The test
applied to the HST data delineates between extended objects and point
sources only, and it is possible that some of the point sources are
Galactic stars.  Therefore this is a {\it lower limit} on the
contamination in the sample.

\subsubsection{Spectroscopic Data}

Another way to distinguish GCs around a galaxy from background or
foreground objects is to measure their radial velocities with
spectroscopy.  Sharples \et (1998) used a wide-field imaging
spectrograph on the 4.2-m Herschel Telescope to obtain radial
velocities for 55 GC candidates around NGC~4472. 47 of these were
confirmed as GCs, 6 had velocities consistent with their being
Galactic stars, and 2 were likely background galaxies.  Zepf \et
(2000) used a multi-object spectrograph on the Canada-France-Hawaii
Telescope to obtain 120 radial velocities for GC candidates around
NGC~4472, finding that 100 of these were GCs, 7 were foreground
objects, and 13 were background objects.

We took the combined results from the Sharples \et (1998) and Zepf \et
(2000) studies and compared them to our GC candidate list to confirm
that our estimate of the amount of contamination in the GC sample
based on the HST data was correct.  The combined spectroscopic data
set (excluding duplicates) contained 134 GCs and 28 contaminating
objects.  We found that 16 of the 28 contaminating objects (5 stars
and 11 galaxies) were present in our sample of GC candidates.

In order to make a meaningful comparison between the spectroscopic
data and our GC sample, we binned the spectroscopic object list into
radial bins from 1\arcmin\ to 9\arcmin\ (as far out as the data
extended).  For each radial bin, we calculated the contamination
fraction, and the total number of contaminating objects (stars and
galaxies) found in that bin.  We executed the same steps for our GC
sample data: we binned the data and calculated a local contamination
fraction, based on the surface density of contaminating objects from
the HST data (0.181 objects per square arc~minute), and the area
contained in the radial bin region.  We used this contamination
fraction to calculate how many contaminating objects we would have
expected to find in the spectroscopic sample of 150 objects.  Note
that this expected number of contaminating objects includes only
galaxies, since the HST data could not distinguish between GCs and
stars.

Eleven galaxies from the spectroscopic sample were included in our
sample of 1,461 GC candidates.  The expected number of galaxies based
on the estimate from the HST data is 9.  This number is consistent,
within Poisson errors, with the actual number found.  Therefore,
results from two independent methods used to estimate the
contamination in our sample due to background galaxies are consistent
with each other.

The fact that five of the 16 contaminating objects from the
spectroscopic sample found in our GC sample were stars indicates that
to account for foreground objects, we should increase the number
density estimate of contaminating objects by $\sim$50\% (or, more
precisely, by the ratio of 5 stars to 11 galaxies).  This increases
the estimated surface density of contaminating objects from 0.181 per
square arc~minute to 0.264 per square arc~minute.

We wished to check that this estimate of the surface density of
foreground stars was reasonable.  Accordingly, the Bahcall-Soneira
model of the Galactic halo (\cite{bs80}) was used to provide an
estimate of the surface density of foreground objects that might
appear in our GC sample.  Since the GC sample contains a limited
magnitude and color range of objects, we imposed a similar selection
on the star counts output from the model.  A selection using two
colors (\bmv\ and \vmr) was not easily implemented in the model code,
so we used a \bmv\ cut only.  The model predicts that the number
density of stars in the direction of NGC~4472 with $V$ magnitude
between 19.5 and 23.5, and \bmv\ color between 0.55 and 1.0, is 114
stars per square degree.  Our estimate of the number density of
objects in the GC sample, based on the spectroscopic data, is 297
stars per square degree, 2.6 times larger than the model prediction.
For this reason we conclude that we are very likely {\it
overestimating} the amount of stellar contamination that exists in our
GC sample.

\subsubsection{Estimate of the Overall Contamination in the GC
Sample}
\label{section:overall contam}
Using two independent methods (HST data and ground-based spectroscopic
data) to estimate the amount of contamination in our GC sample due to
background galaxies, we find that there are approximately 0.181
galaxies per square arc minute in our sample.  We also used two
methods (the spectroscopic data, and Bahcall-Soneira model of the
Galaxy) to estimate the contamination from foreground stars, and
conclude that an upper limit for the number density of foreground
objects in our sample is 0.083 stars per square arc~minute.  Thus the
total surface density of contaminating objects in our sample is
$_<\atop{^\sim}$ 0.264 objects per square arc~minute.  We make use of
this estimate in calculations of the radial distribution and total
number of GCs around NGC~4472, described in
Section~\ref{section:results}.

\subsection{Globular Cluster Luminosity Function}
\label{section:GCLF}

In order to calculate the total number of GCs around NGC~4472, it is
necessary to quantify what fraction of the total GC population (as
characterized by the intrinsic Globular Cluster Luminosity Function
(GCLF) for the galaxy) was detected given the magnitude limits of the
images.  Corrections for magnitude incompleteness in all three
filters, and for contamination from foreground and background objects,
must first be applied to the observed GCLF.  Then the corrected GCLF
can be fit to a GCLF representing the intrinsic one for the galaxy,
and the fraction of area covered by the observed GCLF calculated.  The
intrinsic GCLF is traditionally parameterized by a Gaussian function
of a given peak magnitude and dispersion.

A number of studies of NGC~4472's GCLF have been published in recent
years.  For example, Lee \et (1998) used ground-based data to
determine the GCLF in the Washington $T_1$ band.  Converting their
result to $V$ and applying the value of $A_V$ obtained from the
Schlegel \et (1998) maps yields a GCLF peak at $m^0_V$ =
23.77$\pm$0.10 with dispersion $\sigma$ = 1.3.  Their GCLF
determination came with a number of caveats; for example, it was
affected by contamination from background galaxies at the faint end of
their observed luminosity function.  Puzia \et (1999) used archival
HST data to investigate whether the GCLF turnover varies in different
regions of the galaxy, and in the individual (red and blue) GC
populations.  For the total population, they find the GCLF turnover at
$m^{0}_V$ = 23.76$\pm$0.20.  Lee \& Kim (2000) used the same HST data
to study the GCLF and color distribution of GCs within 4\arcmin\ of
the center of NGC~4472.  They found $m^0_V$ = 23.50$\pm$0.16 and
$\sigma$ = 1.19$\pm$0.09.  The HST observations used for Puzia et
al. and Lee \& Kim reached $\sim$1 magnitude past the GCLF peak.  By
far the best-determined GCLF for an elliptical galaxy in Virgo is that
published by Whitmore \et (1995) using HST observations of M87.  They
observed 1,032 GCs in M87 down to a limiting magnitude of $V$ = 26,
more than 2 magnitudes beyond the GCLF turnover.  Because their data
reached so far past the peak, their determination of both the location
of the peak and, especially, the dispersion of the GCLF are the most
robust of these studies.  Whitmore et al. found $m^{0}_{V}$ =
$23.72\pm0.06$ and $\sigma$ = $1.40\pm0.06$.  We adopt this as the
best estimate of NGC~4472's intrinsic GCLF.  Note that the Whitmore
\et GCLF is consistent within the errors with all the others
mentioned, with the exception that the Lee \& Kim (2000) dispersion,
$1.19\pm0.09$, is significantly more narrow. We attempted to fit our
observed GCLF to the Lee \& Kim (2000) GCLF but found that our data
are more consistent with a GCLF with a larger dispersion.

We derived an observed GCLF by binning the $V$ magnitudes of 1,279 GC
candidates with $V$ $\leq$ 23.5 and radial distance $<$ 16\arcmin\
into 0.4-magnitude-wide bins.  Both the magnitude limit and the radial
cut were imposed to minimize the amount of contamination in the
sample.  (The number density of GCs falls off with increasing radius,
but the number density of contaminating objects is roughly constant;
therefore the fraction of contaminating objects in the sample
increases with distance from the galaxy center.)  The number of GCs
in each bin was multiplied by a correction factor to account for the
fractional contamination present in the sample.

To correct the observed GCLF for magnitude incompleteness, our
completeness curves (see Section~\ref{section:completeness}), were
used to calculate the completeness in each filter for each $V$
magnitude bin.  The completeness in the $V$ filter was calculated by
simply interpolating from the $V$ completeness curve.  Calculating the
completeness for the other two filters was more complicated.  The $B$
completeness that should be applied to a given $V$ bin varies because
the \bmv\ colors of the GCs in the $V$ bin vary over the \bmv\ range
of the sample.  For each of five \bmv\ colors (0.55 to 0.95, at
0.1-magnitude intervals), the $B$ magnitude corresponding to the given
$V$ bin was calculated (i.e., $B$ = (\bmv) + $V$). The completeness
for each of the five $B$ magnitudes was interpolated from the $B$
completeness curve.  The average of these five completeness values was
adopted as the $B$ completeness for that particular $V$ bin.  This
process was repeated for all $V$ bins.  A similar series of steps was
performed to calculate the $R$ completeness for each $V$ bin; a range
of five colors was again used (\vmr\ = 0.34 to 0.6, in steps of 0.065
magnitude) to find the average $R$ completeness corresponding to a
given $V$ bin.  The completenesses in the three filters were then
convolved to calculate a total fractional completeness for each $V$
magnitude bin.  The number of GCs in each bin was divided by this
total completeness to produce the corrected GCLF.  The observed and
corrected GCLF's are shown in Figure~\ref{fig:GCLF}; the solid line
marks the observed GCLF and the dashed line the corrected function.
The constraint that the GC candidates be detected in all three
filters, coupled with the lack of depth of the $B$ image, means that
the observed $V$ GCLF of the GC candidates covers less of the
intrinsic GCLF than desired (see Section~\ref{section:completeness}.

Next, a least-squares fit was performed in order to fit the corrected
GCLF to the GCLF from Whitmore \et (1995).  Bins with completeness
$<$50\% were excluded from the fit. The normalization of the Whitmore
\et GCLF was varied so as to minimize $\chi^2$. The fractional area of
the normalized Whitmore \et GCLF covered by our observed (uncorrected)
GCLF was calculated.  This value, 0.236, gives the fraction of GCs we
detected from the total population of the GC system of NGC~4472.

\section{Results}
\label{section:results}

\subsection{Radial Distribution}
\label{section:profile}
We used the sample of 1,461 GC candidates with $V$ $\leq$ 23.5 to
investigate the radial distribution of GCs around NGC~4472.  First,
the GC candidates were assigned to 1\arcmin-wide bins from 1\arcmin\
to 23\arcmin\ according to their projected radial distances from the
center of the galaxy.  An effective area was calculated for each
annular region.  The effective area equals the area of the annulus,
minus the area of regions excluded from the DAOFIND search for GCs,
and the area of any part of the annulus that extends past the edge of
the Mosaic image. The correction to effective area is very small for
those annuli with central radius $\leq$ 17\arcmin.

To correct for contamination, a subtractive correction (equal to the
number density of contaminating objects times the effective area of
the annulus) was applied to the number of GCs in each radial bin.  A
small multiplicative correction (amounting to multiplying the total
number by 1.04) was then applied to the number of GCs, to account for
faint point sources missing from the GC list because they were
incorrectly excluded in the extended source cut (see
Section~\ref{section:HST data}).  The total number of GCs in each bin
was corrected for incomplete coverage of the GCLF, calculated in
Section~\ref{section:GCLF}.  Finally, the surface density of GCs and
its associated error (calculated from the Poisson error of the number
of GCs, and taking into account errors on the contamination correction
and the correction for GC candidates mistakenly excluded during the
extended source cut) was calculated for each bin, to produce the
radial profile shown in Figure~\ref{fig:profile} and tabulated in
Table~\ref{table:profile}. The table lists the central radius of each
annular region, the number density of GCs within that region, and the
fraction of the annulus for which we have data.

Radial profiles of elliptical galaxy GC systems have traditionally
been fit with power laws or de~Vaucouleurs law profiles, either of
which has provided an adequate fit over the previously-observed radial
ranges, except for the inner 1$-$2 kpc.  Fitting a single power law to
the radial profile yields: log $\sigma_{GC}$ = (1.91$\pm$0.03) $-$
(1.28$\pm$0.04) log $r$, where $\sigma_{GC}$ is the surface density of
GCs and $r$ is in arc minutes.  The single power law does {\it not}
provide a good fit over the entire radial range, however; the surface
density falls well below the power law at large radius.  A
de~Vaucouleurs profile, shown as a dotted line in the bottom panel of
Figure~\ref{fig:profile}, provides a somewhat better fit:
log~$\sigma_{GC}$ = (3.38$\pm$0.08) $-$ (1.56$\pm$0.05) $r^{1/4}$.
The de~Vaucouleurs fit also seems to slightly overestimate the surface
density at large radius, but in this case at least the best-fit line
intercepts nearly all of the error bars in the profile.

The surface density of GCs in the corrected profile is consistent with
zero in the last three bins (21\arcmin\ to 23\arcmin), and is negative
in the 23\arcmin\ bin.  These outer bins have the largest correction
for missing area, since portions of the outer annuli extend beyond the
boundaries of the Mosaic images.  Thus it is not clear whether we have
actually seen the entire extent of the GC system.  On the other hand,
the data do indicate that the distribution of clusters does {\it not}
continue as a power law to radii beyond 23\arcmin\, and the observed
data fall slightly below the de~Vaucouleurs profile (which falls off
more rapidly than a power law) at those large radii.  So while we may
not have seen a clear ``edge'' to the GC system, we do seem to have
seen the point at which the GC system begins to taper off.  Still,
additional data --- with good spatial coverage beyond 23\arcmin ---
must be obtained in order for this result to be confirmed.

\subsection{Color Distribution}

As discussed in Section~\ref{section:completeness}, the $B$-band
Mosaic image from our data set has a 50\% completeness limit that is
slightly brighter than the $V$ image, which means it is much more
shallow than intended.  As a result we are more sensitive to detecting
blue GCs than red ones.  The effect that the shallowness of the $B$
image has on the GC sample is apparent in Figure~\ref{fig:color mag},
which shows the color-magnitude diagram for the 1,465 GC candidates.
The detection limit for these objects becomes brighter as \bmr\ gets
larger (i.e., as the objects become redder).  In order to say
something about the color distribution of the GCs at large radius
around NGC~4472, we needed to define a sub-sample from our GC list
that was equally complete in red and blue clusters.  To do so, we
noted that the reddest object in the total sample had \bmr\ = 1.9.  By
requiring that all clusters with \bmr\ $\leq$ 1.9 must be detected,
one can cut the sample at a given $B$ completeness level (e.g., 50\%)
and use that $B$ magnitude and \bmr\ color to define at what $R$
magnitude the sample must be cut.  For example, the GC sample is 90\%
complete at $B$ = 22.75, so to include the entire red GC population,
the GC candidate list must be cut at $R$ = 20.85.  Doing this yields a
sample of 366 sources that we will refer to as the 90\% sample; their
color distribution is plotted in Figure~\ref{fig:color distn}. The
distribution is clearly bimodal, with peaks at \bmr\ $\sim$ 1.10 and
1.35, and a gap at \bmr\ = 1.23.  If we simply split the color
distribution at the location of the apparent gap, we find that there
are 146 and 220 GCs in the red and blue samples, respectively.  This
yields a ratio of red to blue GCs of $N_{RGC}$/$N_{BGC}$ = 0.66.

To quantify the level of bimodality in the observed color
distribution, we applied the KMM algorithm (\cite{abz94}) to the 90\%
sample.  KMM tests whether a user-specified number of Gaussian
functions provides a better fit to a given distribution than a single
Gaussian, and evaluates the improvement in the fit for the
multi-component model compared to the single-component model.  We find
that for the 90\% sample, the hypothesis that the color distribution
is drawn from a unimodal parent population is rejected with greater
than 99.99\% confidence.  KMM also locates the peaks of the
subpopulations and provides an estimate of the relative proportion of
objects in each subpopulation.  KMM can be run using either a constant
dispersion for the Gaussian functions, or allowing the dispersions to
vary freely (although the second option must admittedly be used with
some caution).

The ratio of the red to blue subpopulations in the 90\% sample
scatters around the value we estimated above using a simple split of
the populations at the location of the gap.  KMM yields a lower ratio
if a fixed dispersion is used, and a higher one if the dispersions are
allowed to vary.  Based on the results from our ``simple'' split and
the KMM test, we estimate the ratio of red to blue clusters to be
approximately 0.7, although we note that this number may be uncertain
by up to 0.2.  In any case, the ratio $N_{RGC}$/$N_{BGC}$ is likely to
be less than one; the implications of this result will be discussed in
Section~\ref{section:models}, when we compare our results with the
predictions of galaxy formation models.

There are too few GC candidates in the 90\% sample to produce robust
radial profiles for the red and blue GC populations individually.
Instead we took the total list of 1,465 GC candidates and separated it
at the location of the gap in the 90\% sample, \bmr\ = 1.23.  Radial
profiles were generated for each sample and fit with power laws; the
slope for the red sample was $-1.34\pm0.08$, and for the blue sample,
$-1.26\pm0.06$.  So although the power law fit for the red clusters is
slightly more steep, the difference is not significant.

An alternative method for comparing the radial distributions of the
two populations is to simply plot the colors of the GC candidates
versus their radial distances.  This plot, made using the 90\% sample,
is shown in Figure~\ref{fig:color radius}.  Lee \et (1998) studied the
GC system of NGC~4472 to a radius of 7\arcmin ($\sim$32 kpc) and found
a strong radial gradient in the color distribution of the total GC
population, in the sense that the GC system becomes bluer with
increasing radius.  They found little, if any, gradient in the
individual blue and red populations, and so concluded that the
gradient in the total population is due to the varying radial mixture
of the two populations --- the red clusters are more centrally
concentrated than the blue clusters.  Figure~\ref{fig:color radius}
shows an apparent overdensity of blue clusters in the region between
$\sim$4\arcmin\ and 8\arcmin, which could give rise to the same type
of color gradient.  When we fit a line to the inner portion
($<$8\arcmin) of the \bmr\ versus radius plot, we find a slightly
negative slope: $\Delta$(\bmr)/$\Delta$$r$ = $-0.010\pm0.007$.  Thus
we find similarly to Lee \et (1998) that the GC system as a whole gets
slightly bluer with increasing radius --- due to the increasing ratio
of blue to red clusters --- in this inner region.

Beyond $\sim$8\arcmin, however, the overdensity of blue GCs
disappears and the overall color distribution becomes more uniform.
The gradient caused by the changing ratio of blue clusters to red
clusters appears to flatten out as a result.  This is borne out by a
fit to the color distribution over the entire radial range of the
data, to 22\arcmin.  The slope for the total population over the whole
radial range is consistent with zero: $\Delta$(\bmr)/$\Delta$$r$ =
$0.000\pm0.002$.

It is possible that the observed lack of a gradient is caused by the
increasing fraction of contaminating objects at larger radii.  Because
there are fewer GCs as the distance from the galaxy center increases,
but the number density of contaminating objects is roughly constant,
the fraction of contaminating objects in the sample increases with
radius.  If for some reason the contaminating objects are
preferentially red, then a larger fraction of them in the outer
regions could cause the measured overall gradient to be zero. 

In order to translate our observed color gradients to metallicity
gradients, we used the Harris catalog of Galactic GCs
(\cite{harris96}) to derive a relationship between \bmr\ and [Fe/H].
We used this relationship to convert \bmr\ to [Fe/H] and then fit
lines to [Fe/H] versus the log of the projected radial distance for
GCs in the inner, outer, and full radial range.  The gradients
($\Delta$[Fe/H]/$\Delta$log($r$)) are $-0.08\pm0.09$ for the full
range, $-0.27\pm0.17$ for the inner 8\arcmin, and $+0.76\pm0.34$ for
the region from 8\arcmin\ to 23\arcmin.  Our metallicity gradient for
the inner radial range falls between that found by Puzia \et (1999),
who surveyed the inner 4.2\arcmin\ of the GC system of NGC~4472 and
obtained $\Delta$[Fe/H]/$\Delta$log($r$) = $-0.15\pm0.02$, and the
Geisler \et (1996) gradient, which was $-0.41\pm0.03$.

\subsection{Specific Frequency}
\label{section:spec freq}

The number of globular clusters in a galaxy can be characterized by
the specific frequency ($S_N$), the number of GCs normalized by the
galaxy luminosity.  Specific frequency is defined as:
\begin{equation}
{S_N \equiv {N_{GC}}10^{+0.4({M_V}+15)}}
\end{equation}
(\cite{harris81}).  It is instructive to calculate both the global
$S_N$ value for a given galaxy, and to investigate the local specific
frequency, or how $S_N$ varies with radius within the galaxy.

\subsubsection{Local Specific Frequency}
\label{section:local S}
Calculating the local specific frequency requires knowledge of the
radial variation of the galaxy light, so that the number of GCs at
each radius can be normalized by the galaxy luminosity at that same
location.  The light distribution of NGC~4472 has been measured in a
number of previous studies.  For example, King (1978) used
photographic data to measure the galaxy's surface brightness out to a
radius of $\sim$17\arcmin.  Fitting a de~Vaucouleurs law to King's
data yields $\mu_V$ = (13.76$\pm$0.14) + (2.52$\pm$0.05)$~r^{1/4}$.
Caon \et (1994) combined data from CCD images and photographic plates
to measure NGC~4472's light distribution out to $r$ $\sim$ 22\arcmin.
The values for the effective radius ($r_e$) and corresponding
effective surface brightness ($\mu_e$) from Caon et al., when
converted from $B$ to $V$ using a mean \bmv\ value for NGC~4472 of
0.95, yield a de~Vaucouleurs law of the form $\mu_V$ = 14.80 +
2.15$~r^{1/4}$.  Most recently, Kim \et (2000) used CCD images to
perform surface photometry of NGC~4472 in the Washington $C$ and $T_1$
bands, with radial coverage to $\sim$9\arcmin.  They found that the
surface brightness distribution in the outer regions of NGC~4472
(beyond $r$=7\arcsec) can be fit approximately with a de~Vaucouleurs
law of the form $\mu_V$ = 13.62 + (2.52$\pm$0.07)$~r^{1/4}$.  (Here
the $T_1$ intercept given in their paper has been converted to a
$V$-band value, using the conversion given in Geisler (1996).)

To measure the galaxy light from our data, we used the IRAF task
ELLIPSE (in the STSDAS package) to perform surface photometry on the
$V$-band Mosaic image of NGC~4472.  We used a bad pixel mask to
exclude regions of the image containing bright stars and other objects
that would interfere with the galaxy photometry.  We measured the sky
background using a median algorithm in an annulur region with inner
radius 16.5\arcmin\ and outer radius 17.5\arcmin, and subtracted the
measured sky level from the image.  The galaxy light was then measured
in elliptical regions with semi-major axes from 47\arcsec\ to
620\arcsec.  Fitting a de~Vaucouleurs profile to the resultant surface
brightness profile yields $\mu_V$ = (14.23$\pm$0.08) +
(2.30$\pm$0.02)$~r^{1/4}$.  Here, $r$ is equal to $\sqrt{ab}$, the
mean of the semi-major and semi-minor axes.  We adopt this as the
galaxy profile for the $S_N$ versus radius calculation.  Note that our
values for the de~Vaucouleurs law coefficients fall between the values
found by Caon \et (1994) and Kim \et (2000).

Figure~\ref{fig:local S} shows the variation of the specific frequency
with radius for NGC~4472.  The data are given in
Table~\ref{table:local S}. The filled triangles in the figure are
local $S_N$ values calculated using our GC data and surface brightness
profile for NGC~4472.  Error bars include Poisson errors on the number
of GCs at each radius, and errors on the corrections we applied to
account for contamination and objects mistakenly excluded during the
extended-source cut (see Section \ref{section:profile}). The solid
line marks the expected rise of $S_N$ if one uses the de~Vaucouleurs
law fit to the radial distribution of GCs to calculate $S_N$ at each
radius.  Values of $S_N$ from the data rise approximately linearly
with radius out to $\sim$10\arcmin.  Beyond that point, the values
scatter substantially above and below the solid line out to
$\sim$16\arcmin.  At larger radii, all of the data points fall below
the line, similarly to the way they do in the radial profile shown in
Figure~\ref{fig:profile}.

Lee \et (1998) studied NGC~4472's GC system using ground-based $C$ and
$T_1$ images and investigated the run of $S_N$ in NGC~4472 out to
$\sim$7\arcmin.  Their values for local $S_N$ are substantially larger
than we find for that radial range. This is due in part to the fact
that we are using a smaller distance modulus (31.12, versus their
31.2) and a more shallow galaxy surface brightness profile, both of
which result in a more luminous galaxy (and therefore a smaller
specific frequency) at each radius.  However, even when we use their
galaxy profile (published in Kim \et (2000)) and their distance
modulus to calculate $S_N$ from our GC distribution, we calculate
$S_N$ values anywhere from $\sim$20 to 80 percent smaller than their
published values.

There were significant differences in the way the GC sample was
selected in Lee \et (1998) compared to the steps we used.  They used a
single color (\cmt) to select GCs from the sample of point sources
they detected around NGC~4472, and were not able to apply image
classification analysis to their sample to eliminate extended sources.
They estimated the amount of contamination in their sample by taking a
number density of contaminating objects around NGC~4472 from a
photographic study by Harris (1986) and estimating what fraction of
those objects would be included after their \cmt\ cut.  For our study,
GCs were selected using image classification and two-color
photometry, and archival HST data and published spectroscopic
observations were used to accurately quantify the amount of
contamination.  Consequently a possible explanation for the
discrepancy between our $S_N$ values and theirs is that their sample
contained a larger fraction of contaminating objects, resulting in
larger $S_N$ values at each radius.

To investigate this possibility, we created a sample of GC candidates
from our data by taking the original sample of 4,329 sources found in
all three (\bvr) Mosaic images and applying a single color cut to the
sample.  We used a \bmv\ cut, which is admittedly less effective for
selecting GCs than the \cmt\ cut used by Lee \et (1998).  As in the
Lee \et study, we did not apply an extended source cut.  The result
was a list of 2,720 GC candidates.  Using this list of sources, we
executed the steps necessary to correct the total number of clusters
for incomplete coverage of the GCLF (i.e., fitting our observed GCLF
to the Whitmore \et (1995) GCLF and calculating the total area under
the Whitmore \et GCLF curve covered by our data).  To correct for
contamination, we doubled the number density of contaminating objects
estimated for our data in Section~\ref{section:overall contam}.
Finally, we calculated local $S_N$ versus radius, using the same
galaxy profile and distance modulus used by Lee et al.  The result
closely matches theirs; our points overlap the points from Lee \et
within the error bars.  Therefore by using less stringent selection
criteria to select our GC sample and underestimating the contamination
level, we are able to reproduce the higher local $S_N$ values found by
Lee \et We believe that our GC sample is much better constrained than
the sample used by Lee et al., and that our local specific frequency
values are consequently more accurate.

We investigated the accuracy of our local $S_N$ values further using a
GC sample identified by Puzia \et (1999) from archival HST images of
NGC~4472.  The data from Puzia \et (1999) are presumably less
contaminated than ground-based data because HST's high spatial
resolution makes it easier to remove background objects from the GC
sample.  These data should therefore provide a useful test of whether
our local $S_N$ values are uniformly too low.  We used a list of 705
GCs provided by T. Puzia and the information given in Puzia \et
(1999) to create a radial profile, correct it for incompleteness, and
fit it with a de~Vaucouleurs law.  The best fit was log~$\sigma_{GC}$
= (3.03$\pm$0.10) $-$ (1.37$\pm$0.09) $r^{1/4}$.  We then calculated
local $S_N$ versus radius for this radial profile and the best-fit
galaxy surface brightness profile from our Mosaic data.  The resultant
$S_N$ values are slightly {\it lower} than a few of our values, but
match most of them within the error bars.  We interpret this as
further evidence that our local $S_N$ values are not underestimated.

Our findings concerning the local specific frequency of NGC~4472 are
relevant to results published in a recent paper by McLaughlin (1999).
A long-standing observational result is that GC systems of ellipticals
are often more spatially extended than the halo light of the galaxy.
This implies that, for some reason, bound clusters were sometimes more
likely to form at large galactic radius, from gas with comparatively
low densities and pressures.  McLaughlin explores this issue for
NGC~4472 using a composite GC radial profile, created by combining
published CCD and photographic observations of the galaxy's GC system.
Using the composite profile and the galaxy profile from Caon \et
(1994), McLaughlin finds that the ratio of the mass surface density of
the GC system to that of the stellar component of the galaxy is
constant at radii beyond $\sim$15 kpc.  This in turn implies that the
local specific frequency, which is directly proportional to the ratio
of the mass density of GCs to stars, is roughly constant throughout
NGC~4472's outer halo.  We find, on the contrary, that NGC~4472's GC
system is substantially more extended than the stellar component; the
effective radius of the globular cluster profile ($r_e$) is several
times that of the galaxy light.  As a result the local specific
frequency, rather than being constant at large radius, increases
linearly out to at least $\sim$10\arcmin\ ($\sim$49 kpc), and perhaps
somewhat beyond that.  

\subsubsection{Global Specific Frequency}

To calculate the global $S_N$ for the GC system of NGC~4472, we begin
by calculating the total number of GCs in the galaxy.  We can
calculate a total number, corrected for magnitude incompleteness, lack
of spatial coverage, and contamination from non-GCs, by taking the
best-fit de~Vaucouleurs law to our radial profile (calculated as
described in Section~\ref{section:profile}, and shown in
Figure~\ref{fig:profile}) and integrating it from $r$ = 0 to some
outer radius.  The choice of outer radius is not obvious.  To
calculate a total number of GCs for the entire system, one could
integrate the de~Vaucouleurs profile to $r$ $=$ $\infty$; doing so
gives $\sim$11,100 GCs.  Integrating the galaxy profile obtained from
our Mosaic data from $r$ = 0 to $\infty$ yields a total $V$ magnitude
for NGC~4472 of $V^T$ = 8.01.
(Note that this is $\sim$0.4 magnitudes brighter than the $V$
magnitude given in de~Vaucouleurs \et (1991).)
This corresponds to an absolute magnitude of $M_V^T$ = $-$23.11 for
the distance modulus determined by Whitmore \et (1995), $m - M$ =
31.12.  Using these values yields a specific frequency $S_N$ = 6.4.

The calculation above, however, almost certainly overestimates the
total number of GCs.  We argue that based on the appearance of the
radial profile in Figure~\ref{fig:profile}, integrating to infinity is
not a valid choice, since the data points in the profile fall below
the de~Vaucouleurs fit at large radii, and the surface density of GCs
is consistent with zero in the last three bins.  A better choice would
be to integrate the de~Vaucouleurs profile to the limit of the data
(23\arcmin, or $\sim$110 kpc) and calculate an $S_N$ value for
NGC~4472 {\it out to that radius}.  Integrating the de~Vaucouleurs
profile and the galaxy profile to 23\arcmin\ yields 5,870 clusters and
$M_V^T$ = $-$23.05 for the galaxy.  This results in a specific
frequency $S_N$ = 3.6.

To calculate an error on the global specific frequency, we need to
quantify the amount of error associated with each of the major sources
of uncertainty in $S_N$.  One contribution is the uncertainty in the
total number of GCs.  Assuming Poisson statistics, we can calculate
uncertainties in both the total number of GCs observed, and the number
of contaminating objects, and add these in quadrature.  Doing this, we
find that the uncertainty on the total number of GCs leads to an
uncertainty in $S_N$ of $\sim$0.1.  A second source of error comes
from the uncertainty in the galaxy profile and thus the amount of
galaxy light out to 23\arcmin.  As discussed in
Section~\ref{section:local S}, other groups have found steeper light
profiles for NGC~4472, which leads to a slightly fainter total
magnitude at 23\arcmin.  For example, if we use the galaxy profile
from Kim \et (2000) to calculate $S_N$, the galaxy magnitude $M_V^T$
is $-$22.91 at 23\arcmin, resulting in an $S_N$ value of 4.0.
(Applying the Caon \et (1994) profile, on the other hand, produces no
change in $S_N$.)  We estimate that the uncertainty in $S_N$ due to
the galaxy profile is $\sim$0.5.  Finally, there is some error
associated with the correction made to account for our incomplete
coverage of the GCLF.  The correction varied by $\sim$10\% depending
on which observed GCLF was used for the fit to the intrinsic GCLF
(i.e., exactly how the observed GCLF was corrected for contamination,
and whether a radial cut was applied to the GC sample).  This
translates to an uncertainty in $S_N$ of $\sim$0.4.  Combining the
three main sources of uncertainty in quadrature yields a final value
of $S_N$ for NGC~4472 of 3.6$\pm$0.6.

Our value of $S_N$ is smaller than values found by authors of previous
studies of NGC~4472, who typically found $S_N$ values in the range
$\sim$5$-$6.  Studies based on CCD data had much less spatial coverage
than our data, usually extending out to only a few to several arc
minutes.  In such cases, authors extrapolated their observed radial
profiles out to some radius.  For example, Lee \et (1998) extrapolated
their GC profile out to only 10\arcmin, and calculated a specific
frequency $S_N$ = 4.7$\pm$0.6 (for $V^T$ = 8.4 and $(m - M)$ = 31.2).
Photographic studies had spatial coverage comparable to ours but had
higher levels (often more than an order of magnitude higher) of
contamination.  A photographic study of NGC~4472's GC system by Harris
(1986, 1991) finds $S_N$ = 5.0$\pm$1.4 (for $V^T$ = 8.4, $(m - M)$ =
31.3, and integrating to 25\arcmin).  The $S_N$ values from Lee \et
(1998) and Harris (1986, 1991) are both larger than ours, but are
consistent within the errors.

We consider the value of $S_N$ = 6.4 --- calculated by integrating our
profile from $r$ $=$ 0 to infinity --- an absolute upper limit on the
total $S_N$ for NGC~4472.  Previous studies that found $S_N$ values of
5$-$6 likely had more contamination from foreground and background
objects than was accounted for.  We suggest that $S_N$ = 3.6$\pm$0.6,
calculated by integrating our profile to $r$ $=$ 23\arcmin, is a more
accurate estimate of the true value for NGC~4472, since our observed
radial profile indicates that it is unlikely that another $\sim$5,000
GCs exist at radii beyond 23\arcmin.  

\section{Comparison to Model Predictions}
\label{section:models}

Although we have just begun this project and therefore cannot draw
general conclusions about the GC systems of ellipticals and spirals,
we can use the results for NGC~4472 to test a few specific predictions
of models for galaxy formation.

Here we consider four different galaxy formation models.  The first is
the monolithic collapse scenario (e.g., \cite{larson75},
\cite{carlberg84}, \cite{ay87}), in which elliptical formation is
modeled as the collapse of an isolated massive gas cloud, or
``protogalaxy'' at high redshift.  Such scenarios would produce a GC
system with a color distribution that has a smooth shape and single
peak directly related to the metallicity of the population.  A second
type of scenario is that in which ellipticals are formed by mergers,
e.g., by the merger of two or more spirals (e.g., \cite{toomre77},
\cite{az92}, \cite{za93}).  The simple merger scenario suggested by
Ashman \& Zepf (1992; hereafter AZ92) involves the merger of disk
galaxies, and yields two GC populations: a younger,
spatially-concentrated, metal-rich system of clusters formed during
the merger, and a spatially-extended, metal-poor system that comes
from the halos of the disk galaxies.  A model by Forbes, Brodie, \&
Grillmair (1997; hereafter FBG97) combines elements of both collapse
and merger-type scenarios, since it involves dissipational collapse as
well as accretion of surrounding galaxies.  In the FBG97 scenario,
ellipticals form their GCs in distinct star formation phases.  The
first phase forms a system of metal-poor GCs, and the second phase
(which occurs after the gas in the galaxy has been self-enriched)
gives rise to a population of more metal-rich GCs.  FBG97 suggest
that the outer regions of giant cluster ellipticals contain a
population of metal-poor GCs that have been captured either through
tidal stripping or accretion of smaller galaxies.  The most recent
proposed formation model that makes predictions for GC system
properties is that put forth by C\^ot\'e, Marzke, \& West (1998;
hereafter CMW98).  CMW98 propose that giant ellipticals form by
dissipational collapse, during which a metal-rich population of GCs
is formed.  The population of bluer, metal-poor GCs are subsequently
captured from other galaxies through mergers or tidal stripping.

\noindent {\it Radial profile and local specific frequency.}  Because
our fractional areal coverage drops beyond $\sim$17\arcmin\ from the
center of NGC~4472, we can by no means state unequivocally that we
have observed the entire extent of the GC system.  Our observed radial
profile does indicate, however, that {\it it is unlikely that large
numbers of GCs reside at radial distances beyond $\sim$23\arcmin}.
Models such as FBG97 and CMW98, that invoke tidal stripping as the
mechanism by which GC systems of giant cluster ellipticals are formed,
require that a large reservoir of GCs be present in the outer regions
of galaxies.  We have observed NGC~4472 out to $\sim$110 kpc and do
not find significant numbers of GCs at those radial
distances. Indeed, the local specific frequency of GCs remains below
15 out to 23\arcmin.

\noindent {\it Ratio of red to blue clusters.}  Like previous authors
(e.g., \cite{geisleretal96}, \cite{puzia99}), we find that NGC~4472's
GC system has a bimodal color distribution.  A simple monolithic
collapse model cannot explain the presence of two GC populations, so
it appears that this particular galaxy, at least, has a more
complicated formation history.  We also find, as did Geisler \et
(1996), that the ratio of red (metal-rich) to blue (metal-poor)
clusters in NGC~4472 is less than one.  The AZ92 model, which
attempted to explain why elliptical galaxies have specific frequencies
$_>\atop{^\sim}$2 times those of spirals, suggests that the merger of
two or more spiral galaxies creates a population of GCs comparable in
size to the number originally contained in the spirals.  This has the
result of increasing $S_N$ by the required amount, and requires that
$N_{RGC}/N_{BGC}$ $_>\atop{^\sim}$ 1.  This is not observed in
NGC~4472, in disagreement with the AZ92 model.  On the other hand, the
total number of GCs in ellipticals compared to spirals is still not
well-known.  It is possible that the differences between the specific
frequencies of typical ellipticals and spirals are smaller than
previously thought, in which case the requirement from AZ92 that an
equal number of GC's be created in the merger would no longer hold.
These issues are explored further in the discussion below about global
specific frequency.

\noindent {\it Color gradient.}  All of the models we are considering
can explain why the blue population of GCs in ellipticals would be
more spatially extended than the red GCs.  Lee \et (1998) observed a
color gradient in the total GC population of NGC~4472 out to
$\sim$7\arcmin\ and showed that this was caused by the increasing
ratio of blue to red GCs at large radius.  We observe a similar
(although smaller) color gradient in the inner portion of the
galaxy, but the gradient vanishes for the total population over the
entire radial extent of the data.  The flattening of the color
gradient is not predicted by, nor consistent with, any of the proposed
models.  In fact, if the zero-gradient result is real, it raises some
interesting questions about how the metal-rich GCs were formed at
large radii.

\noindent {\it Global specific frequency.}  We find a smaller global
specific frequency for NGC~4472 than previous studies, although our
value agrees with some previous values within the errors.  We maintain
that our GC sample is significantly less contaminated compared to past
studies, due to our combined techniques of observing through multiple
filters and using image analysis to eliminate extended objects.
Moreover, using spectroscopic data and HST images, we have accurately
quantified the amount of contamination in the sample so that we can
then correct for it.  Consequently we are better able to constrain the
total number of GCs in NGC~4472.

Accurately establishing the total numbers of GCs in both spirals and
ellipticals is of central importance in the study of extragalactic GC
systems.  Although ellipticals are generally thought to have specific
frequencies at least twice as large as those for spirals, the
quantitative difference between spirals and ellipticals is still
uncertain.  The reasons for this are that few spirals have been
studied in detail, and the uncertainties in individual measurements of
specific frequency for both spirals and ellipticals are large.  Past
studies have typically used photographic plates or small-area CCD's,
and thus had low sensitivity and/or poor spatial coverage, and also
had higher levels of contamination.  Accurately-determined specific
frequencies provide an important test of formation models, since all
the models seek to explain the differences in specific frequencies for
disk galaxies versus ellipticals.

One of the main goals of our overall study is to better determine the
total numbers of GCs for each of a large sample of spiral galaxies, in
order to test the prediction made by AZ92 that the mass-normalized
number of metal-poor GCs around ellipticals and spirals should be the
same.  Even with the somewhat smaller total number of GCs we have
observed in NGC~4472, it appears that there may be too many blue GCs
in this galaxy to have originated in the halos of typical spirals.  A
useful parameter for comparing the total numbers of GCs in galaxies of
widely varying masses is the $T$ parameter, the number of GCs per
stellar galaxy mass (\cite{za93}).  $T$ is defined as:
\begin{equation}
T \equiv \frac{N_{GC}}{M_G/10^9\ {\rm M_{\sun}}}
\end{equation}
where $M_G$ is the mass of the host galaxy.  The total number of
clusters in NGC~4472 out to 23\arcmin\ is $\sim$5,900; using this and
the ratio of red to blue GCs, we calculate that about 3,500 of the
GCs are blue and metal-poor.  In the AZ92 picture, these are the
clusters that originated in the halos of the progenitor spirals that
merged to form NGC~4472.  Assuming a mass-to-light ratio $M/L_V$ = 10
for ellipticals (cf.~\cite{fg79}), we obtain $T$ = 2.6 for the
metal-poor GCs in NGC~4472.

If we calculate $T$ for the {\it entire} GC system of the two most
well-studied spirals, the Milky Way and M31, we obtain $T$ values of
approximately 1 and 2, respectively.  (Here we have used $N_{GC}$ =
180 and $M/L_V$ = 5 for the Galaxy, and $N_{GC}$ = 450 and $M/L_V$ = 6
for M31.)  If we assume that the Milky Way and M31 are typical
spirals, then it would appear that even with our scaled-down total
number of GCs, {\it the AZ92 scenario cannot account for the blue
cluster population observed in giant ellipticals like NGC~4472.}

The assumption that the Milky Way and M31 are typical may not be a
valid one.  There are a few spiral galaxies that have been found by
previous studies to have $N_{GC}$ $\sim$900$-$1,000 (see summary in
Appendix of Ashman \& Zepf (1998)), yielding $T$ values
$_>\atop{^\sim}$3.  However, in some of these cases less than 10\% of
the estimated number of GCs were actually observed, making the $T$
values highly uncertain. Still other spirals have been found to have
$T$ $<$1, and the total number of GCs around a ``typical'' spiral is
very much an open question.  Based on our experience with NGC~4472, it
is possible that when we analyze the data for the spiral galaxies in
our sample, we will (because of our techniques for reducing
contamination) find smaller specific frequencies for them as well.

In conclusion, our observations are consistent with some aspects of
each of the models considered, and inconsistent with at least one
prediction or assumption of each model.  AZ92 make specific
predictions concerning the characteristics of the GC systems of
individual galaxies, so their model was perhaps the simplest to test
using only one galaxy from our larger sample.  It seems likely from
this and other studies of NGC~4472 that the gaseous merger scenario
{\it as proposed by AZ92} is perhaps overly simplistic and that more
complicated formation processes must be invoked in order to explain
exactly how the GC system of NGC~4472 originated.  Our study of
isolated spirals and ellipticals in various environments is designed
to help answer some of the outstanding questions concerning
extragalactic GC systems, and address some of the more detailed
predictions of FBG97 and CMW98, both of which make testable
predictions concerning the characteristics of the GC systems of
ellipticals in general.

\section{Summary}
\label{section:summary}

We have recently begun a study of the GC systems of a large sample of
elliptical and spiral galaxies, in order to test the predictions of
models for elliptical galaxy formation.  Below we summarize our
findings for the first elliptical observed for this study, the giant
Virgo galaxy NGC~4472:

1.  A de~Vaucouleurs law of the form log~$\sigma_{GC}$ =
(3.38$\pm$0.08) $-$ (1.56$\pm$0.05) $r^{1/4}$ provides a fairly good
fit to the radial profile of NGC~4472's GC system, out to
$\sim$16$-$17\arcmin.  At larger radii, the observed profile falls
slightly below the de~Vaucouleurs law, and the number density of GCs
is consistent with zero in the region from 21\arcmin\ to 23\arcmin.

2. The bimodal color distribution in NGC~4472's GC system observed by
previous authors is apparent in our \bmr\ distribution.  The overall
ratio of red to blue GCs is $\sim$0.7.

3. For GCs located within 8\arcmin\ of the galaxy center, we find a
small negative color gradient: the GC system becomes bluer with
increasing radius, due to the increasing ratio of blue to red
clusters.  The gradient disappears, however, when the entire GC
population out to 22\arcmin\ is taken into account; {\it we find no
color gradient in the total GC population.}

4.  The local specific frequency rises approximately linearly with
radius out to $\sim$10\arcmin\ ($\sim$50 kpc).  At larger radii, the
local $S_N$ values scatter substantially above and below the values
predicted by the de~Vaucouleurs law fit to the radial profile.  In the
region from 17$-$23\arcmin, the observed local $S_N$ appears to turn
over and begin to decrease.  Our values of local $S_N$ are
significantly lower than those found by Lee \et (1998) using
ground-based data, but agree fairly well with values calculated from
an HST study by Puzia \et (1999). We do {\it not} find, as suggested
by McLaughlin (1999), that local $S_N$ is constant at all radii beyond
$\sim$15 kpc.

5.  There are $\sim$5,900 clusters in NGC~4472's GC system out to
23\arcmin, yielding a global specific frequency within that radius of
3.6$\pm$0.6.

6. We have compared our results with the predictions of four different
galaxy formation models, and find that the results are inconsistent
with at least one aspect of each model.  Our observed radial profile
indicates it is unlikely that a large population of GCs resides at
radii beyond $\sim$23\arcmin, challenging formation models that
require a substantial reservoir of GCs in the far outer regions of
galaxies.  The ratio of red to blue clusters for NGC~4472 is at odds
with a prediction of the simple merger scenario that the numbers of
metal-rich and metal-poor GCs should be comparable.  The lack of a
color gradient in the total GC population is not explained nor
predicted by any of the models considered.  Finally, although our
calculated total number of GCs in NGC~4472 is smaller than previous
estimates, it appears that there are still slightly more metal-poor
GCs per unit mass in NGC~4472 than can be accounted for by merging
the halo GC populations of spiral galaxies like the Milky Way and M31.



\acknowledgments

K.L.R. gratefully acknowledges financial support from a NASA Graduate
Student Researchers Fellowship during the preparation of this paper.
S.E.Z. acknowledges support from the Hellman Family Fellowship and
NASA Long-Term Space Astrophysics grant NAG5-9651.  We are grateful to
Thomas Puzia for providing an electronic version of the GC candidate
list from Puzia \et (1999), and to Charles Liu for providing
electronic versions of the template galaxy spectra used for the
color-color plane analysis.  We thank the staff at Kitt Peak National
Observatory for their assistance during our 4-meter observing run.  We
also thank John Salzer, Doug Geisler, Thomas Puzia, and the anonymous
referee for useful comments that have improved the quality of the
paper.  This research has made use of the NASA/IPAC Extragalactic
Database (NED), which is operated by the Jet Propulsion Laboratory,
California Institute of Technology, under contract with the National
Aeronautics and Space Administration.


\clearpage

%
%

\clearpage

\begin{figure}
\plotone{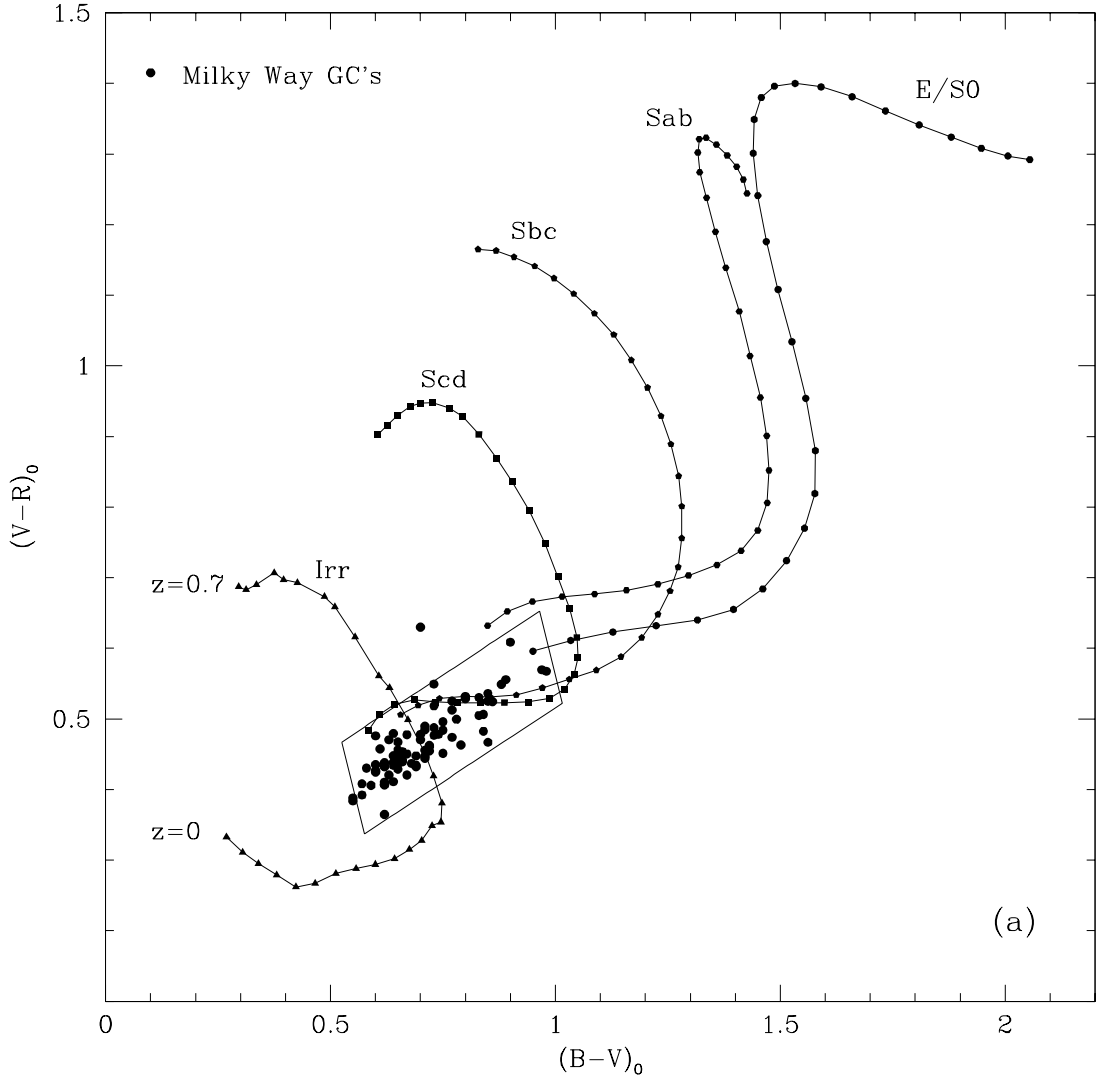}
\caption{{\bf (a) Milky Way GCs in the BVR color-color plane.}  The
rectangle marks the region where GCs would be expected to appear in
this color-color plane.  Milky Way GCs from the Harris catalog
(\cite{harris96}) are shown as filled circles.  The positions of
galaxies of different morphological types appear as ``tracks'' the
galaxies would follow with increasing redshift (see
Section~\ref{section:color cut} for details).  {{\bf (b) Color
selection.}  Open squares mark the positions of the 3,458 GC
candidates that passed the extended source cut in the \bvr\
color-color plane.  Filled circles mark the 1,465 sources whose colors
and associated photometric errors put them within 2-sigma of the \vmr\
versus \bmv\ relation for GCs.  For reference, the galaxy tracks that
appear in Figure~2$a$ are also shown here.  The reddening vector
appears in the upper left corner, and its length corresponds to $A_V$
$=$ 1 magnitude.}}
\end{figure}

\begin{figure}
\plotone{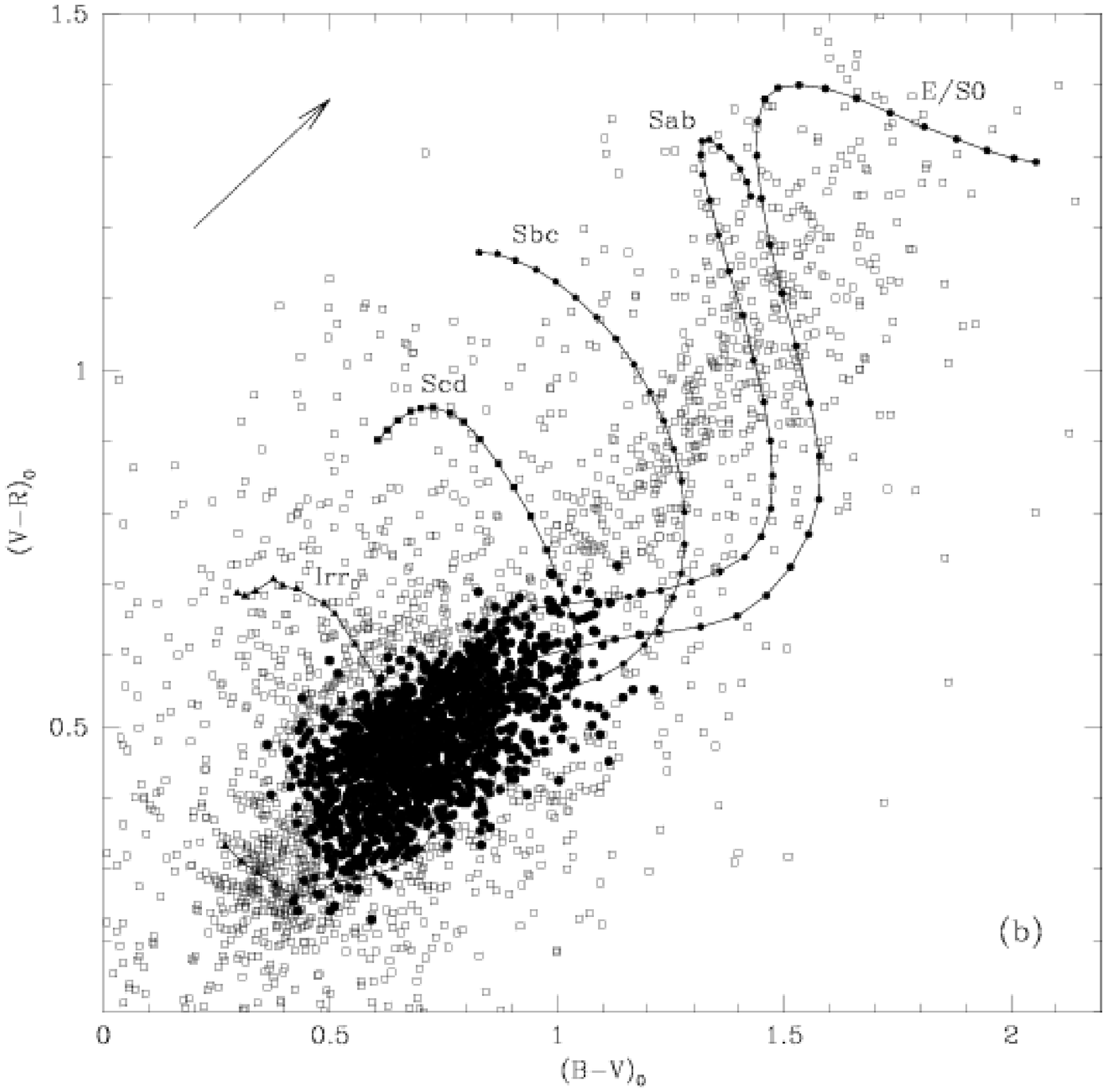}
\end{figure}

\begin{figure}
\plotone{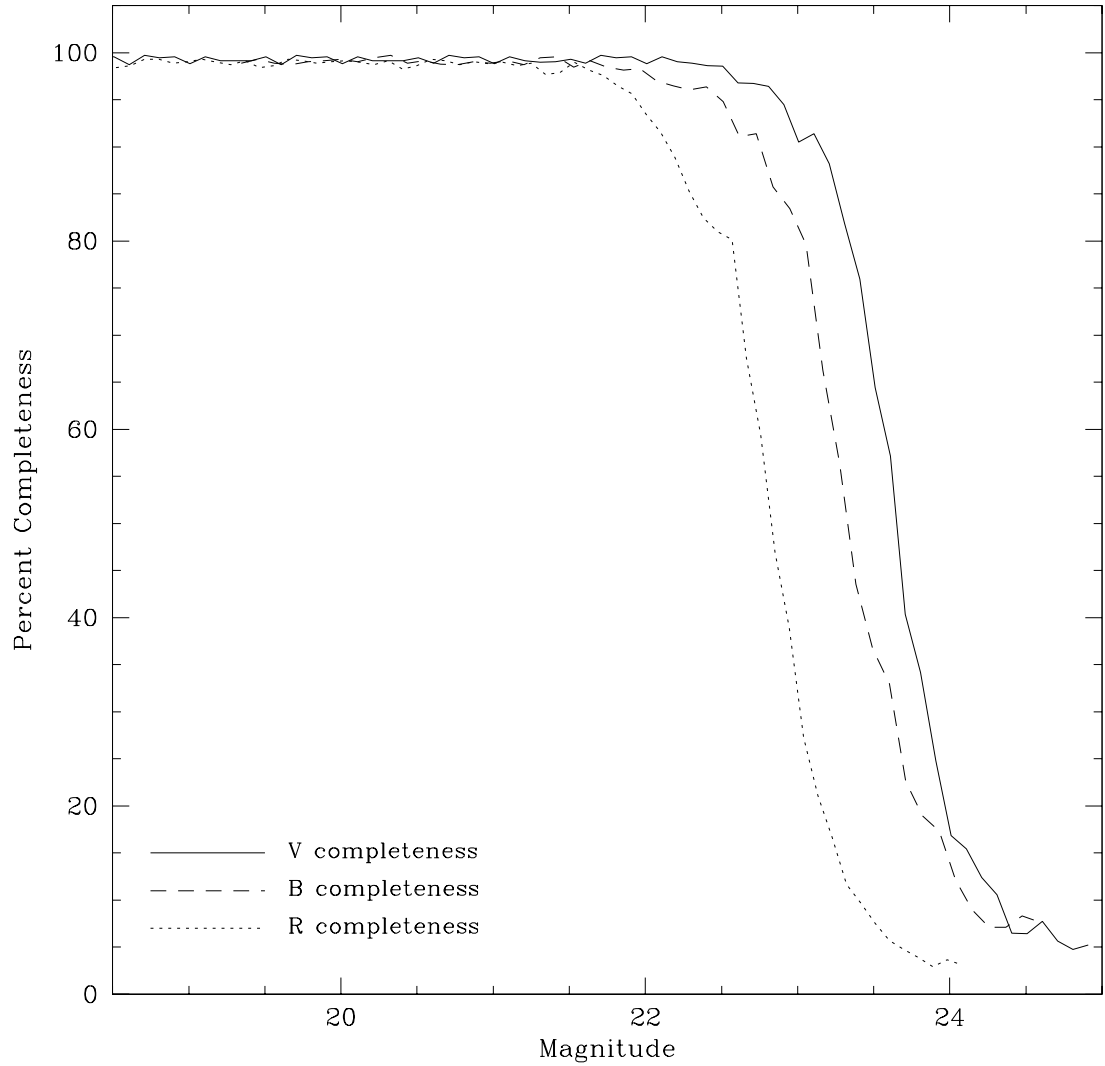} 
\caption{{\bf Completeness as a function of magnitude.}  The detection
limits for GCs for our images were determined using artificial star
tests, as described in Section~\ref{section:completeness}.  The
resultant completeness curves for each filter are shown here.
\protect\label{fig:completeness}}
\end{figure}

\begin{figure}
\plotone{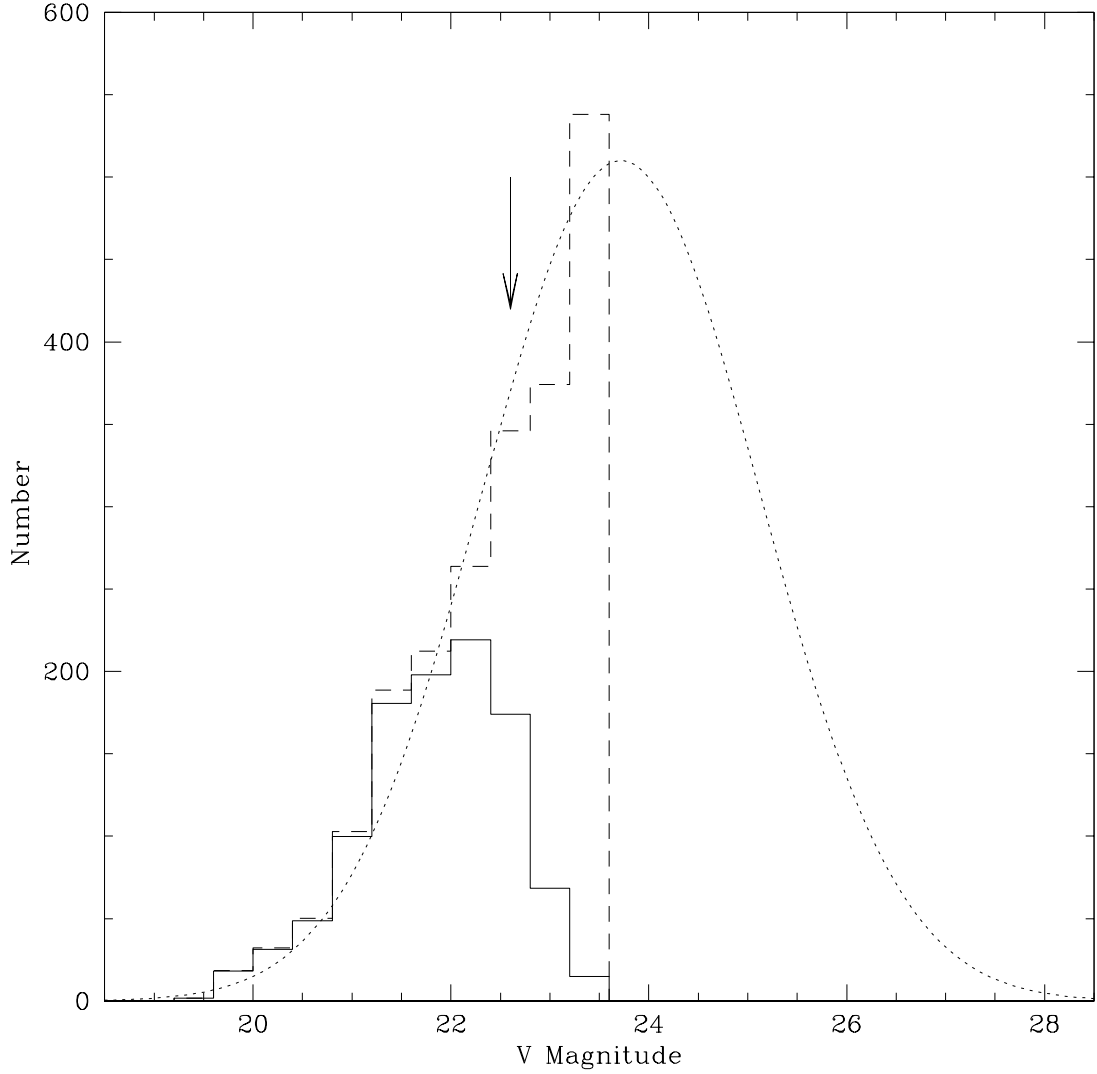} 
\caption{{\bf Globular Cluster Luminosity Function.}  The solid line
marks the observed $V$-band luminosity function for 1,279 GC
candidates with $V$ $\leq$ 23.5, located within 16\arcmin\ of the
center of NGC~4472.  The dashed line shows the same GCLF corrected for
incompleteness in the filter combination ($B$, $V$, and $R$) used to
select the GC candidates.  The arrow points to the bin where the
combined completeness is 50\%.  The M87 GCLF from Whitmore \et (1995),
normalized to our data, is shown as a dotted line.
\protect\label{fig:GCLF}}
\end{figure}

\begin{figure}
\plotone{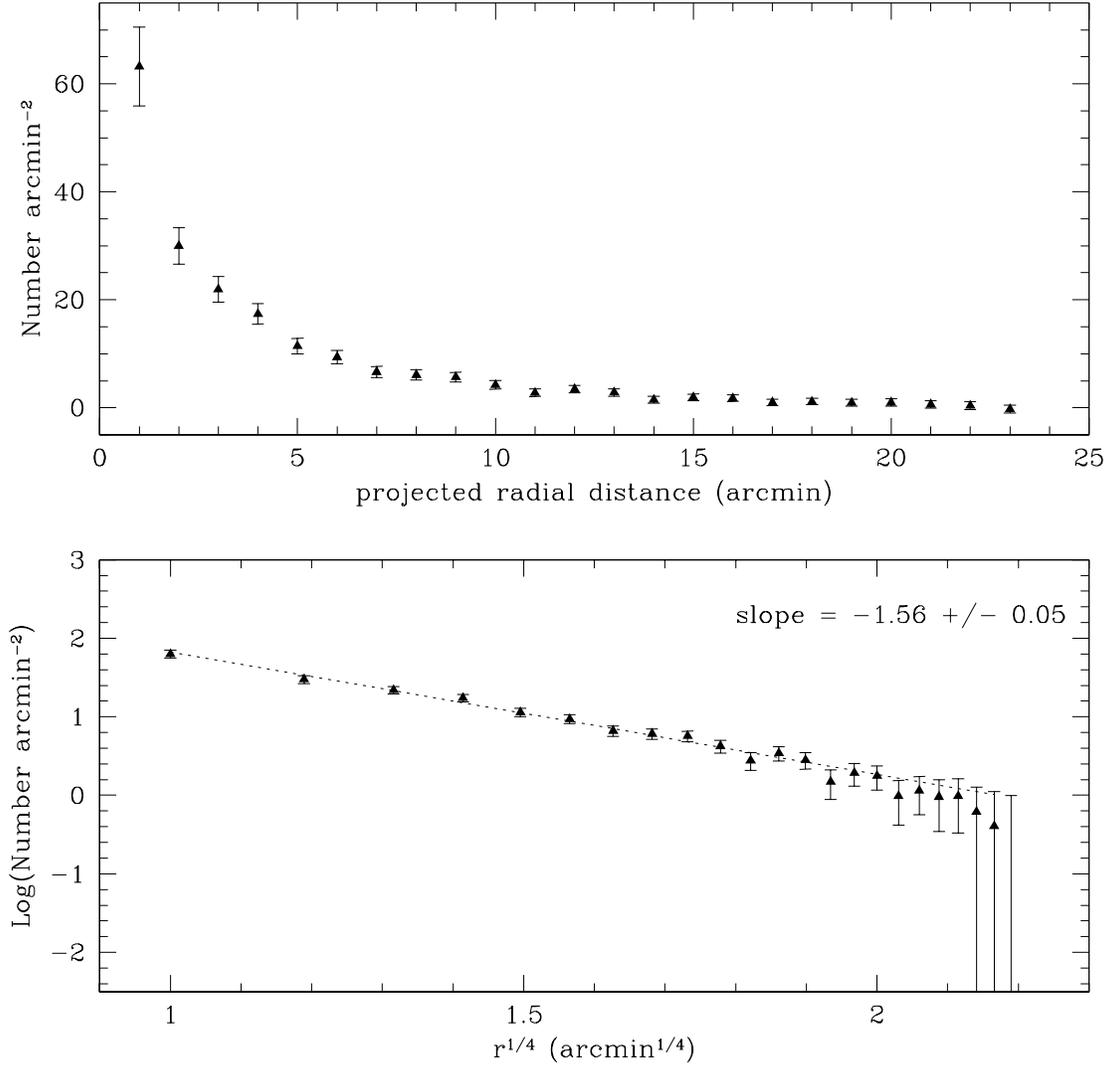} 
\caption{{\bf Radial distribution of GCs.} The upper plot shows the
surface density of GCs in NGC~4472 as a function of projected radial
distance.  The lower plot shows the log of the surface density versus
r$^{1/4}$, and the dotted line is the best-fit de~Vaucouleurs profile.
The profiles have been corrected for contamination, areal coverage,
and magnitude incompleteness, as described in
Section~\ref{section:profile}.  \protect\label{fig:profile}}
\end{figure}

\begin{figure}
\plotone{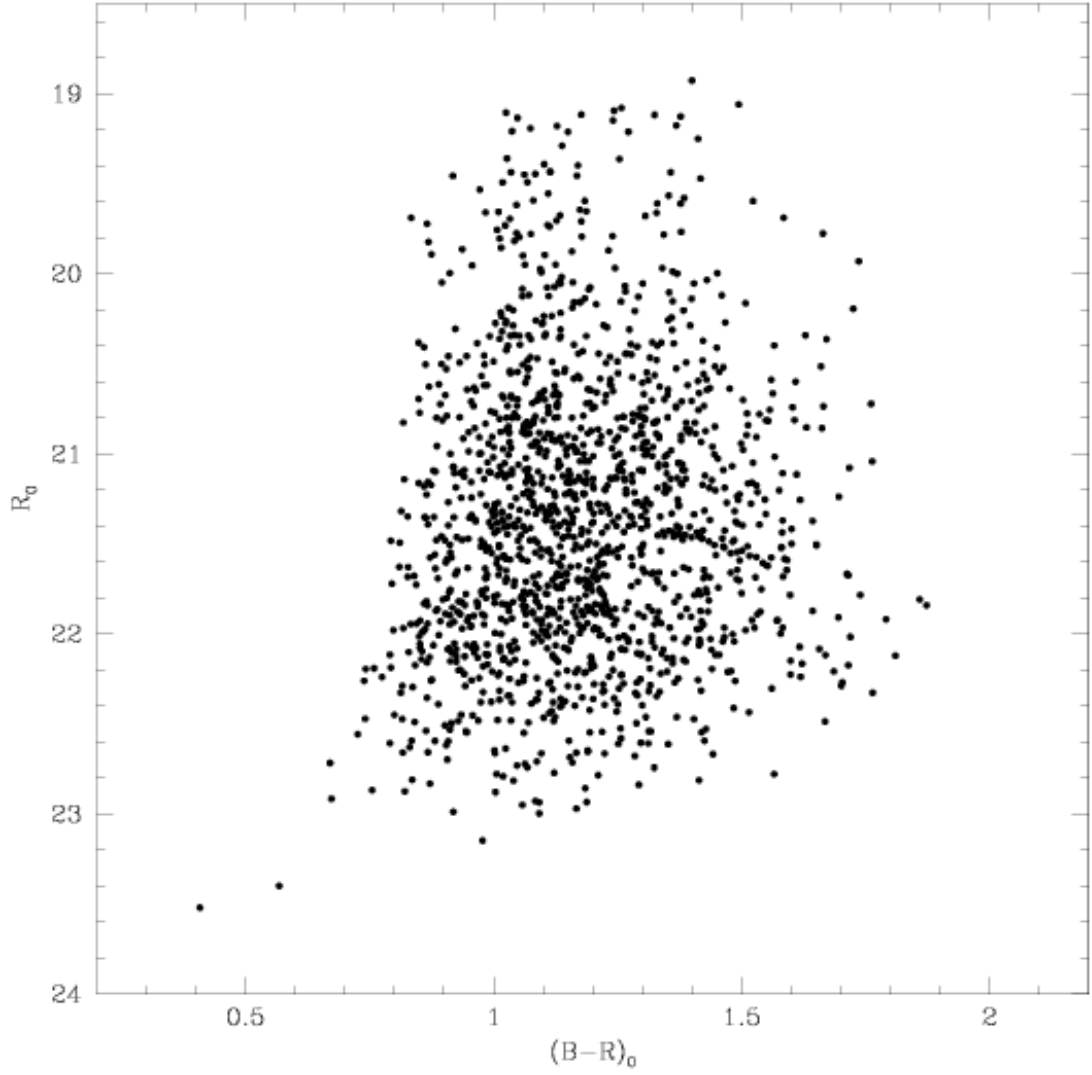} 
\caption{{\bf Color-magnitude diagram for GC candidates.} $R$
magnitudes for 1,465 GC candidates are shown here versus their \bmr\
colors.  The relative shallowness of the $B$ image means that we are
more sensitive to detecting blue GCs than red ones.
\protect\label{fig:color mag}}
\end{figure}

\begin{figure}
\plotone{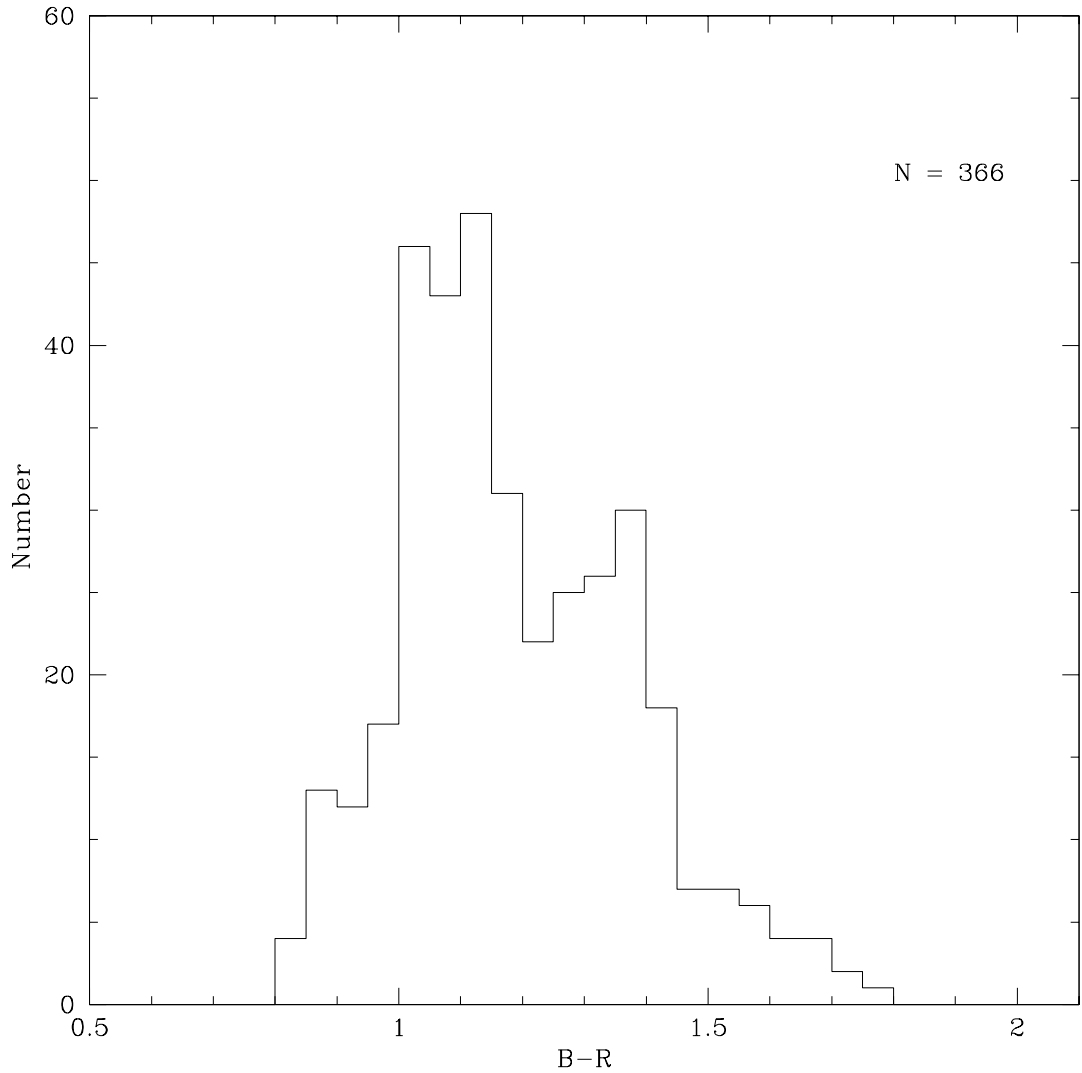} 
\caption{{\bf Color distribution of GC candidates.}  This plot shows
the color distribution of a sub-sample of objects taken from the total
sample of 1,465 GC candidates and selected to be 90\% complete for the
blue population and complete for the red population.  The \bmr\
distribution is bimodal, with peaks around \bmr\ = 1.10 and 1.35, and
a gap at 1.23.  \protect\label{fig:color distn}}
\end{figure}

\begin{figure}
\plotone{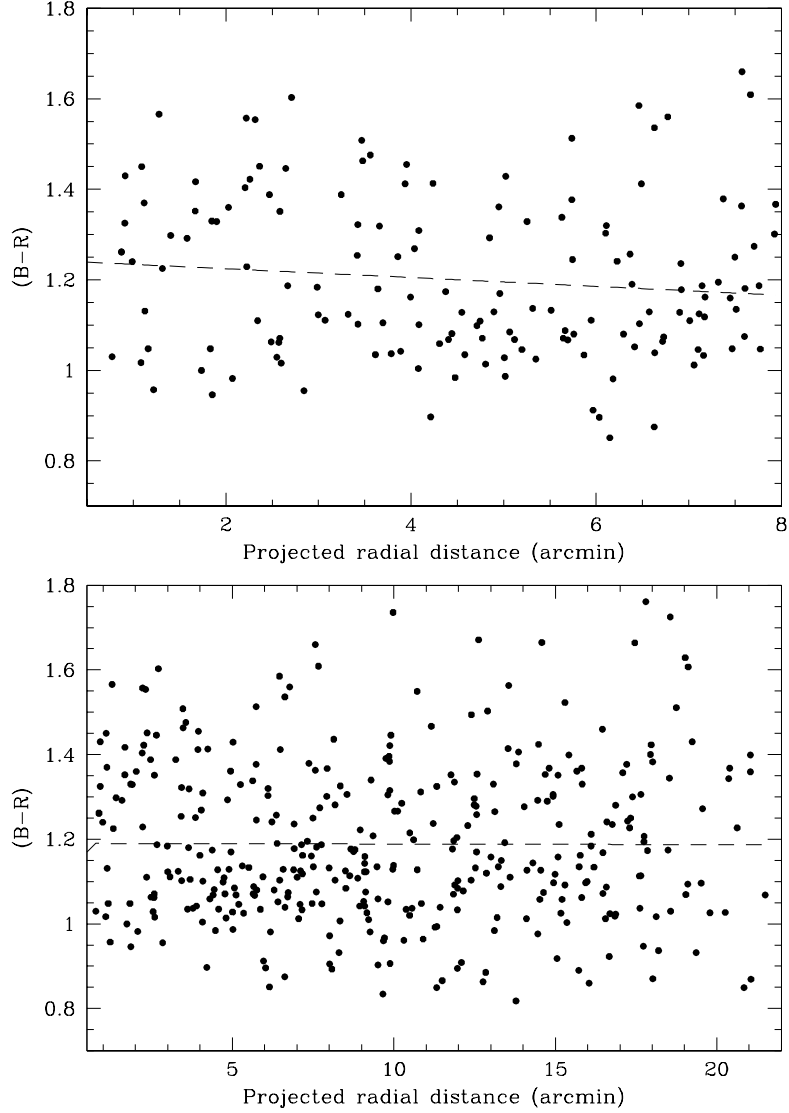} 
\caption{{\bf Color versus projected radial distance for 366 GC
candidates.}  These plots show \bmr\ color versus radius for the 90\%
sample of GC candidates, for the inner 8\arcmin\ of NGC~4472 (upper
plot) and the entire radial region covered by our data (lower plot).
The small gradient seen in the color distribution for the inner region
is due to a slight overdensity of blue clusters in the radial range
4\arcmin\ to 8\arcmin.  The color gradient vanishes in the total
population for the full radial range of the data.  Dashed lines are
best-fit lines to the data shown.  \protect\label{fig:color radius}}
\end{figure}

\begin{figure}
\plotone{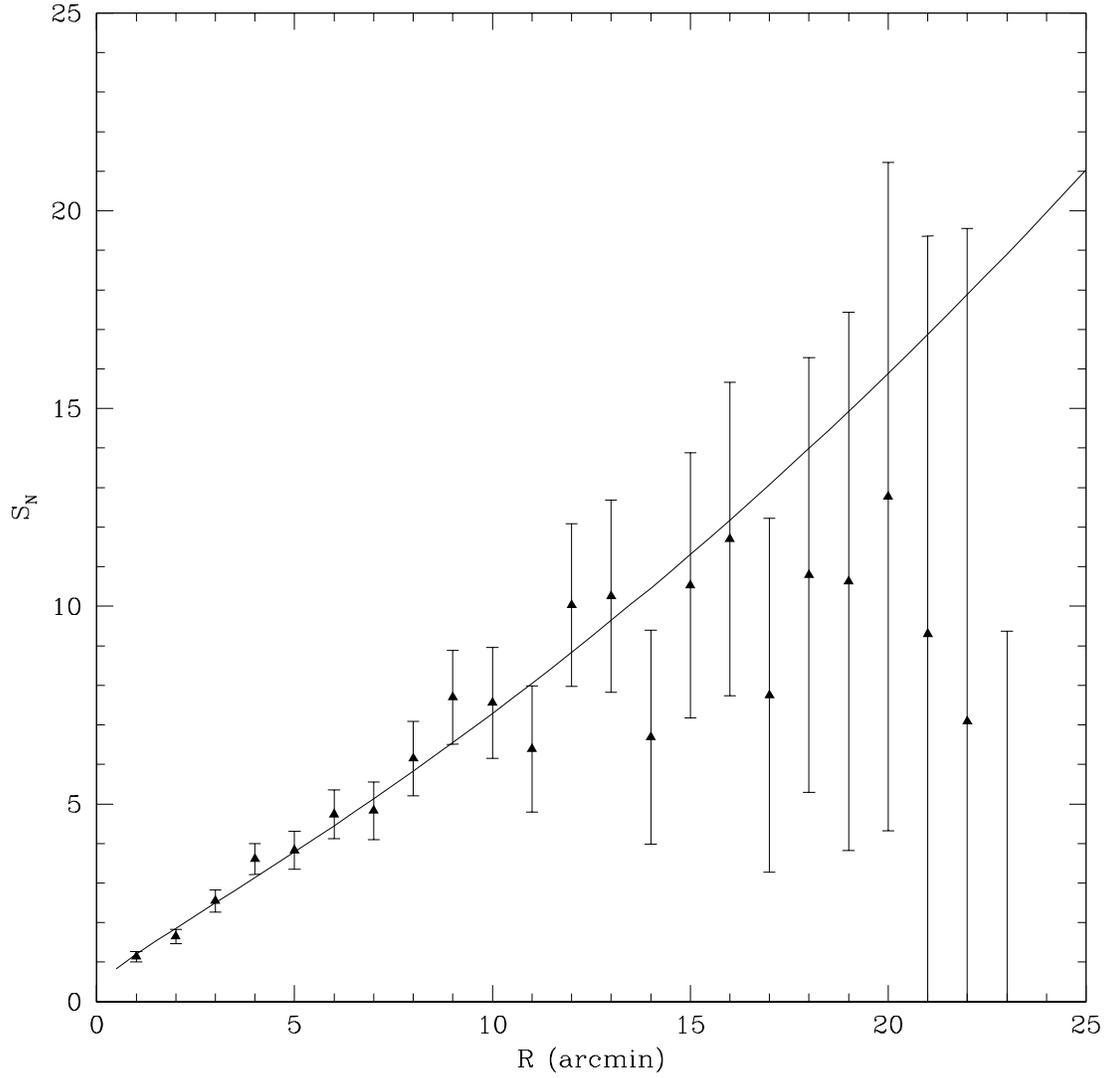} 
\caption{{\bf Local specific frequency of GCs.} This figure shows how
$S_N$ varies with radius in NGC~4472.  Filled triangles are points
calculated using our GC sample and surface brightness profile for
NGC~4472.  The solid line shows the expected $S_N$ based on the
best-fit de~Vaucouleurs law fit to the radial distribution of GCs.
The errors in $S_N$ become larger as fewer GCs are detected within the
area of each annular region.  \protect\label{fig:local S}}
\end{figure}

\clearpage
\renewcommand{\arraystretch}{.6}
\begin{deluxetable}{lcc}
\tablecaption{Aperture Corrections Used for Photometry of GC Candidates}
\tablewidth{200pt}
\tablehead{\colhead{Image} & \colhead{Aperture Correction} & \colhead{}}
\startdata
$B$ & $-$0.2112 $\pm$ 0.0184\nl
$V$ & $-$0.1970 $\pm$ 0.0117\nl
$R$ & $-$0.1376 $\pm$ 0.0176\nl
\enddata
\protect\label{table:apcorr}
\end{deluxetable}

\begin{deluxetable}{crc}
\tablecaption{Corrected Radial Profile for GCs in NGC~4472}
\tablewidth{250pt}
\tablehead{\colhead{Radius} & \colhead{$\sigma$} & \colhead{Fractional coverage}\nl
\colhead{(arcmin)} & \colhead{(arcmin$^{-2}$)} & \colhead{}}
\startdata
1 &  63.23 $\pm$ 7.33 &  0.86\nl   
2 &  29.98 $\pm$ 3.37 &  1.00\nl      
3 &  21.93 $\pm$ 2.40 &  1.00\nl      
4 &  17.39 $\pm$ 1.88 &  1.00\nl      
5 &  11.44 $\pm$ 1.42 &  1.00\nl      
6 &  9.38 $\pm$ 1.22 &  1.00\nl      
7 &  6.64 $\pm$ 1.01 &  1.00\nl      
8 &  6.10 $\pm$ 0.94 &  1.00\nl      
9 &  5.68 $\pm$ 0.88 &  1.00\nl      
10 &  4.24 $\pm$ 0.78 &  1.00\nl      
11 &  2.79 $\pm$ 0.70 &  0.98\nl      
12 &  3.45 $\pm$ 0.71 &  1.00\nl      
13 &  2.82 $\pm$ 0.67 &  1.00\nl      
14 &  1.49 $\pm$ 0.60 &  1.00\nl      
15 &  1.93 $\pm$ 0.61 &  1.00\nl      
16 &  1.77 $\pm$ 0.60 &  1.00\nl      
17 &  0.98 $\pm$ 0.57 &  1.00\nl      
18 &  1.15 $\pm$ 0.59 &  0.81\nl      
19 &  0.96 $\pm$ 0.61 &  0.54\nl      
20 &  0.98 $\pm$ 0.65 &  0.39\nl      
21 &  0.61 $\pm$ 0.66 &  0.28\nl      
22 &  0.40 $\pm$ 0.71 &  0.18\nl      
23 &  $-$0.25 $\pm$ 0.71\tablenotemark{\dag} &  0.10\nl      
\enddata
\tablenotetext{\dag}{A subtractive correction for contamination has been
applied to the surface density of GCs in each bin, resulting in a
negative surface density for the outermost bin.}
\protect\label{table:profile}
\end{deluxetable}

\begin{deluxetable}{cr}
\tablecaption{Radial Variation of Specific Frequency for GCs in NGC~4472}
\tablewidth{200pt}
\tablehead{\colhead{Radius (arcmin)} & \colhead{$S_N$}}
\startdata
 1 &      1.14    $\pm$ 0.13\nl
 2 &      1.65    $\pm$ 0.18\nl
 3 &      2.55    $\pm$ 0.28\nl
 4 &      3.61   $\pm$ 0.39\nl
 5 &      3.83   $\pm$ 0.48\nl
 6 &      4.74   $\pm$ 0.62\nl
 7 &      4.83   $\pm$ 0.73\nl
 8 &      6.15   $\pm$ 0.94\nl
 9 &      7.70   $\pm$  1.19\nl
 10 &     7.56   $\pm$  1.40\nl
 11 &     6.39   $\pm$  1.60\nl
 12 &    10.03   $\pm$  2.06\nl
 13 &    10.25   $\pm$  2.43\nl
 14 &     6.69   $\pm$  2.70\nl
 15 &    10.53   $\pm$  3.35\nl
 16 &    11.70   $\pm$  3.96\nl
 17 &     7.75   $\pm$  4.47\nl
 18 &    10.79   $\pm$  5.50\nl
 19 &    10.63   $\pm$  6.80\nl
 20 &    12.77   $\pm$  8.45\nl
 21 &     9.30   $\pm$ 10.06\nl
 22 &     7.09   $\pm$ 12.46\nl
 23 &    $-$5.06   $\pm$ 14.43\nl
\enddata
\protect\label{table:local S}
\end{deluxetable}

\end{document}